\documentclass[usenatbib]{mn2e}
\usepackage{amsmath}
\usepackage{txfonts}
\usepackage{graphicx}
\usepackage{array}

\newcommand{\be}{\begin{equation}}
\newcommand{\ee}{\end{equation}}
\newcommand{\bea}{\begin{eqnarray}}
\newcommand{\eea}{\end{eqnarray}}

\newcommand{\oiii}{[O\,{\sc iii}]}

\newcommand{\civ}{C\,{\sc iv}}
\newcommand{\ciii}{C\,{\sc iii}}

\newcommand{\lya}{Ly-$\alpha$}
\newcommand{\nv}{N\,{\sc v}}

\newcommand{\co}{CO\,{\sc (1$\rightarrow$0) }}

\newcommand{\hst}{{\sl HST} }
\newcommand{\vla}{{\sl VLA} }

\newcommand{\jvla}{{\sl JVLA } }
\newcommand{\evn}{{\sl EVN} }
\newcommand{\vlba}{{\sl VLBA } }
\newcommand{\merlin}{{\sl MERLIN} }

\newcommand{\mnras}{MNRAS}
\newcommand{\nat}{Nat}
\newcommand{\aj}{AJ}
\newcommand{\apj}{ApJ}
\newcommand{\apjl}{ApJL}
\newcommand{\aap}{A\&A}
\newcommand{\apjs}{ApJS}

\title[Preferential magnification in IRAS\,F10214+4724  \--- III.]{\mbox{The preferentially magnified active nucleus in IRAS\,F10214+4724 \--- III.} VLBI observations of the radio core}

\author[Deane et al.]{R.P. Deane$^{1,2}$\thanks{E-mail: roger.deane@astro.ox.ac.uk}, S. Rawlings$^{1}$, M.A. Garrett$^{3,4}$, I. Heywood$^{1}$, M.J. Jarvis$^{1,5,6}$, 
\newauthor H.-R. Kl\"ockner$^{1,7}$, P.J. Marshall$^{1}$, J.P. McKean$^3$
\vspace*{3pt}\\
\noindent $^1$Astrophysics, Department of Physics, University of Oxford, Keble Road, Oxford, OX1 3RH, UK \\
\noindent $^2$Astrophysics, Cosmology and Gravity Centre, Department of Astronomy, University of Cape Town, Private Bag X3, Rondebosch 7701, South Africa \\
\noindent $^3$ASTRON, The Netherlands Institute for Radio Astronomy, Postbus 2, 7990 AA, Dwingeloo, The Netherlands \\
\noindent $^4$Leiden Observatory, Leiden University, Postbus 9513, 2300 RA Leiden, The Netherlands \\
\noindent $^5$Centre for Astrophysics Research, Science \& Technology Research Institute, University of Hertfordshire, Hatfield, AL10 9AB, UK \\
\noindent $^6$Physics Department, University of the Western Cape, Cape Town, 7535, South Africa \\
\noindent $^7$Max-Planck-Institut f\"ur Radioastronomie, Auf dem H\"ugel 69, 53121 Bonn, Germany }

\begin{document}

\date{Accepted 2013 July 6}

\pagerange{\pageref{firstpage}--\pageref{lastpage}} \pubyear{2013}

\maketitle

\label{firstpage}

\begin{abstract} We report 1.7 GHz Very Long Baseline Interferometry (VLBI) observations of IRAS\,F10214+4724, a lensed $z=2.3$ obscured quasar with prodigious star formation. We detect what we argue to be the obscured active nucleus with an effective angular resolution of $<$~50~pc at $z = 2.3$. The $S_{\rm 1.7} = 210 \, \mu$Jy (9-$\sigma$) detection of this unresolved source is located within the \hst rest-frame ultraviolet/optical arc, however, $\gtrsim$100 milli-arcseconds northward of the arc centre of curvature. This leads to a source plane inversion that places the European VLBI Network detection to within milli-arcseconds of the modelled cusp caustic, resulting in a very large magnification ($\mu \sim$70), over an order of magnitude larger than the \co derived magnification of a spatially resolved \jvla map, using the same lens model. We estimate the quasar bolometric luminosity from a number of independent techniques and with our \mbox{X-ray} modelling find evidence that the AGN may be close to Compton-thick, with an intrinsic bolometric luminosity $\log_{10}(\langle L_{\rm bol,QSO} \rangle/{\rm L}_{\odot}) = 11.34 \pm 0.27$~dex. We make the first black hole mass estimate of IRAS\,F10214+4724 and find $\log_{10}(M_{\rm BH}/\mathrm{M_{\odot}})  =  8.36 \pm 0.56$ which suggests a low black hole accretion rate ($\lambda = \dot{M} / \dot{M}_{\rm Edd} \sim 3\pm^7_2$~percent). We find evidence for a $M_{\rm BH}/M_{\rm spheroid}$ ratio that is 1-2 orders of magnitude larger than that of submillimetre galaxies (SMGs) at $z \sim 2$. At face value, this suggests IRAS\,F10214+4724 has undergone a different evolutionary path compared to SMGs at the same epoch. A primary result of this work is the demonstration that emission regions of differing size and position can undergo significantly different magnification boosts ($> 1$~dex) and therefore distort our view of high-redshift, gravitationally lensed galaxies.
\end{abstract}

\begin{keywords}
gravitational lensing: strong, galaxies: evolution, galaxies: active 
\end{keywords}

\section{Introduction}\label{section_introduction}

Two major empirical results in the last 15 years have greatly increased the need to characterise the black hole accretion history of the Universe, in particular in the redshift range $z \sim 2-3$ where the cosmic star formation and black hole accretion rate densities appear to peak \citep[e.g.][]{Madau1996,Dunlop1990}. The first of these breakthroughs is the finding that super-massive black hole masses are strongly correlated with their host galaxy spheroid luminosity and stellar velocity dispersion \citep{Magorrian1998,Gebhardt2000,Ferrarese2000}. The second is the realisation that most ($\gtrsim$70 percent) black hole accretion is obscured \citep[e.g.][]{Martinez-Sansigre2005,Gilli2007}. Not only is the role of super-massive black holes more fundamental than previously thought, but our view of this fundamental process is largely hidden for the majority of active galaxies. 
This has driven a number of (necessarily multi-wavelength) observation programmes to characterise this population of `obscured' quasars, with a focus on the X-ray and mid-infrared windows \citep[e.g.][]{Polletta2006,Fiore2008}. However, these regions of the electromagnetic spectrum have their challenges and contaminants, predominately due to Compton-thick column densities; the level of torus and/or host galaxy obscuration; and the unknown contribution of star formation to mid-infrared dust heating. In this regard, high resolution radio imaging provides a direct method of disentangling the AGN and star formation components through the measurement of brightness temperature and the radio spectral index. 

This was demonstrated by \citet{Klockner2009} who showed a sample of eleven $z \gtrsim 2$ obscured quasars to be radio intermediate luminosity quasars ($L_{\rm 5 GHz} = 10^{23-24}$~W\,Hz$^{-1}$\,sr$^{-1}$), a challenging observation for VLBI arrays. Until the previous decade, VLBI was an observational technique used almost exclusively to detect bright ($\gtrsim$10~mJy) radio sources which could be self-calibrated to improve the image quality and dynamic range. However, due to a confluence of factors - technical and scientific - VLBI has entered the sub-mJy domain.

In this work, we exploit the unique advantage of VLBI observations to isolate high brightness temperature emission at milli-arcsecond scales and constrain the active core flux and position in a high-redshift, gravitationally lensed obscured quasar. While large-scale VLBI surveys will only be possible with the full {\sl Square Kilometre Array (SKA)}, current studies of individual systems (particularly lensed systems) provide a preview of the science that will be possible with a large sample size in a decade. In addition, a larger sample of unlensed systems from the {\sl Herschel ATLAS} catalogue \citep{Eales2010} have been observed with the \evn (Virdee~et~al., in preparation), a study that is important since it is free from any magnification bias and preferential magnification effects.

This paper is the third in a series on the gravitationally lensed, obscured quasar IRAS~F10214+4724 (IRAS~10214 hereafter). IRAS~10214 is a well-studied galaxy that is often used as an archetype high-redshift ULIRG, largely because of its extensive multi-wavelength coverage and its early discovery in 1991 \citep{Rowan-Robinson1991}. It radiates  $\gtrsim$95~percent of its bolometric luminosity in the infrared, the intrinsic source of which is poorly constrained. IRAS~10214 is classified as a cusp-caustic lens that results in a large arc to the south of the lens galaxy, and a small counter-image to the north of the lens \citep[see][]{Eisenhardt1996}. In \citet[][{\bf D13a} hereafter]{Deane2013a} we derived a new lens model for IRAS~10214 based on a deep \hst\,F814W map and introduced methods to investigate preferential magnification. In Deane~et~al.~(2013b, {\bf D13b} hereafter) we report a spatially resolved \co \jvla observation that, by proxy, provides an estimate of the magnification of the star formation component in IRAS~10214. 
 
 In this work we focus on the AGN properties of IRAS~10214. Previous observations have provided strong arguments that IRAS~10214 hosts an obscured quasar. The primary pieces of evidence supporting this are as follows.
 
 \begin{enumerate}
 
\item The observed optical continuum emission is highly polarised,  implying that this is predominately scattered rest-frame ultraviolet light from a dust-embedded source \citep{Lawrence1993}. 

\item Optical spectra show a wide range in ionisation, supporting the view that the galaxy hosts an active nucleus \citep{Elston1994,Soifer1995}

\item There are clear broad emission lines present in the polarised spectrum which are typical of quasars \citep{Goodrich1996}. 
 
 \end{enumerate}

Despite these strong pieces of evidence, X-ray observations did not confirm the X-ray luminosity ($L_{\rm X}$) expected in IRAS~10214, particularly when compared with the measured \oiii$\lambda$5007 luminosity which is correlated with $L_{\rm X}$ for type I AGN \citep{Alexander2005,Iwasawa2009}. These authors concluded that the X-ray emission associated with IRAS~10214 was either dominated by a dust-enshrouded starburst or that the active nucleus was Compton thick ($\sigma_{\rm T} \gtrsim10^{24} \ \mathrm{cm}^{-2}$). 
 

The obscured nature of this quasar clearly makes it challenging to disentangle its emission from that of the host galaxy and reliably correct for extinction. In light of these challenges, we carried out VLBI observations which are unaffected by dust extinction and able to isolate high brightness temperature emission originating from an active radio core.  Building on the previous multi-wavelength work and incorporating the unique characteristics of VLBI observations, the aims of this paper are threefold: 
 
 \begin{enumerate}

\item Detect the obscured active nucleus in a $z\sim2$ radio quiet quasar and hence estimate its contribution to the total radio flux.

\item Use the position to determine the AGN magnification. This is part of a wider case study of this high-redshift galaxy which also aims to quantify the scale of preferential lensing in this system, that is to say, the level of distortion in the spectral energy distribution (SED) that occurs due to different physical emission regions undergoing different magnification boosts. This is important in demonstrating the level of effective `chromaticity' in strong gravitational lensing (due to the background source properties) which affects the physical interpretations of these systems. 

\item VLBI imaging has the potential to solve a question that arose following the high resolution \hst imaging \citep{Eisenhardt1996,Nguyen1999,Evans1999}: is the UV/optical/NIR arc in this system a single image, or three images that are unresolved with the \hst spatial resolution? The root cause of this debate stems from the two apparent peaks along the \hst\,F814W and \hst\,F437M arcs which could be intrinsic source structure (i.e. clumpy UV emission as is typically observed in high-redshift galaxies), or multiple, partially resolved images. 

\end{enumerate}

This paper is structured as follows: in \S2 and \S3 we describe the \evn observations and present the results. \S4 reviews the ultraviolet polarisation properties of IRAS~10214, while \S5 investigates the source-plane position, magnification and multi-wavelength context of the VLBI-detected source. In \S6 and \S7 we estimate the quasar bolometric luminosity and black hole properties; and close with conclusions in \S8. Throughout this paper we assume a concordance cosmology of $\Omega_{\rm M}$~=~0.27, $\Omega_{\Lambda}$~=~0.73, and $H_0$~=~71~km\,s$^{-1}$\,Mpc$^{-1}$ \citep{Spergel2007}, which yields an angular size scale of 8.3~kpc\,arcsec$^{-1}$ at the redshift of IRAS 10214 ($z$ = 2.2856, \citealt{Ao2008}).

\section{Observations}

\subsection{EVN 1.7 GHz}\label{sec:evnobs}

Observations of IRAS~10214 were made with the \evn at 1.66~GHz on 2 and 3 November 2010. Stations that were used in these observations included Jodrell Bank (76-metre), Westerbork Radio Synthesis Array (WRST, phased array), Effelsberg, Onsala (26-metre), Medicina, Torun, Cambridge, Knockin, as well as the three Russian out-stations: Svetloe, Zelenchuk and Badary. This resulted in a baseline coverage of roughly 100 to 5000 km, corresponding to an angular scale range of 500 to 10~mas at an observing frequency of 1.66~GHz. It is for this reason that VLBI observations apply a brightness temperature filter which enables the unambiguous determination of the active core radio flux - one of the primary aims of this work. Observations were carried out with a 1024~Mbit/s recording rate in the standard continuum observing mode, employing eight 16~MHz sub-bands in each hand of polarisation with 2-bit sampling. The total bandwidth of 128~MHz was centred on 1.65899~GHz. Each sub-band was split into 32 channels of 500~kHz width. The integration period was set to 4~seconds. This results in a time- and bandwidth-smearing limited field-of-view of 0.24 and 0.34~arcmin$^2$ respectively. This is based on a metric that measures the angular displacement from the pointing centre at which a 10~percent loss in the response to a point source is measured. These effective fields of view are larger than the optical/IR and molecular extent of IRAS~10214 by two orders of magnitude ({\bf D13b}).

The total duration of the observations was 18~h, of which roughly 70~percent was spent on the target. The observations were split into two 9~h runs that were performed on consecutive days. Given the IRAS~10214 1.7~GHz flux density of $S_{\rm 1.7} \sim 1$~mJy, as measured in {\bf D13a}, the phase referencing technique was required. Through a separate \evn calibrator search we selected the \mbox{$S_{\rm 1.7} \sim 70$~mJy} source {\sl J}\,1027+474 as the best possible phase calibrator for IRAS~10214 given its small angular separation of $\Delta \theta \sim$~34~arcmin. IRAS~10214 was observed for 8 min every 10~min, alternating with the phase calibrator. There was an additional cycle where the standard \vlba phase calibrator ({\sl J}\,1027+4803) was observed for 4 min every hour. This was done to monitor the applied phase corrections and to check the resultant astrometric accuracy. 4C39.25 was observed as a fringe-finder. Since the observation was split into 2$\times$9~hr tracks; and the \evn has limited north-south coverage; the resulting beam pattern has a large ripple in the north-south direction (evident in Fig.~\ref{fig:evn}). 
 
Preliminary calibration of the {\sl uv}-data set was carried out with the \evn automatic pipeline. This pipeline is written in {\sc parseltongue} \citep{Kettenis2006}, a high-level environment/interface for {\sc aips}, and performs initial fringe-fitting (calibration of delays, rates and phase), and phase and amplitude calibration. The latter are derived from system temperature measurements at individual stations. Following this preliminary calibration, a more detailed, manual calibration is performed in {\sc aips}. This was performed in a cyclic process with detailed data editing, refined delay and rate calibration, as well as phase and amplitude self-calibration on the phase reference sources. Unfortunately, the majority of the observation on the second day was lost due to a combination of circumstances including strong winds, radio frequency interference (RFI), as well as the lack of several stations which did not take part (Effelsberg, Torun).

All imaging was performed with the {\sc aips} {\tt \sc Imagr} task with a natural weighting scheme applied to the {\sl uv-}data. Only after several rounds of the phase and delay calibration cycle (performed on the phase calibrator) did the noise level decrease to \mbox{$\sigma \sim 23~\mu$Jy\,beam$^{-1}$} for a {\sl uv-}range limited between 0 - 5 M$\lambda$. Including longer baselines ($>$5 M$\lambda$) improves the noise marginally (20~percent) however it results in large and small-scale features that decrease the overall fidelity of the IRAS~10214 map. The reason for this decrease in image quality is due to the fact that the phase calibrator is spatially resolved on the longest baselines, leading to poor phase solutions. For this reason we only consider baselines for which the phase calibrator is clearly unresolved (0 - 5 M$\lambda$).

The measured absolute position of the standard \vlba calibrator {\sl J}\,1027+4803 is within $\Delta \theta <$ 0.80 mas of its cataloged position. This was true for the preliminary pipeline calibration and for the manual delay and  self-calibration performed in {\sl AIPS}. The catalogued position uncertainties are ($\Delta$RA = 0.28 mas, $\Delta$Dec = 0.43 mas), sourced from the International Celestial Reference Frame (ICRF-2). These two positions are consistent to within the sub-milliarcseond level, greatly exceeding the level of accuracy required in this work. The integrated flux density of {\sl J}\,1027+4803 ($S_{\rm int} = 149 \pm 2$~mJy) is consistent with that derived by \citet{Helmboldt2007} with their \vlba 5 GHz observations ($S_{\rm int,5GHz} = 149.3 \pm 0.2$~mJy). VLBI resolution {\sl L}-band observations of {\sl J}\,1027+4803 have not been performed before prior to this work, so no direct flux density comparison is possible. 

In addition, the radio coordinate reference frame must be directly compared to the optical reference frame. This has been performed by \citet{Lawrence1993} who compare the optical and radio positions of 20 compact radio sources in the region around IRAS~10214, however, they do not explicitly state the wavelength at which this comparison is done. They find no significant mean difference, however they note the error on the mean difference is 0.2~arcsec which we take as the systematic uncertainty in the radio-optical reference frame alignment. Greater detail is provided in \citet{Lawrence1993}, however, since we use the same data, and our calibrators (as well as the 8~GHz centroids) are consistent, we assume there is an equivalent astrometric matching between radio (1.7 and 8 GHz) and optical reference frames.

\section{Results from 1.7 GH\lowercase{z} VLBI map}\label{sec:resultsVLBI}

\begin{figure}
\includegraphics[width=0.47\textwidth]{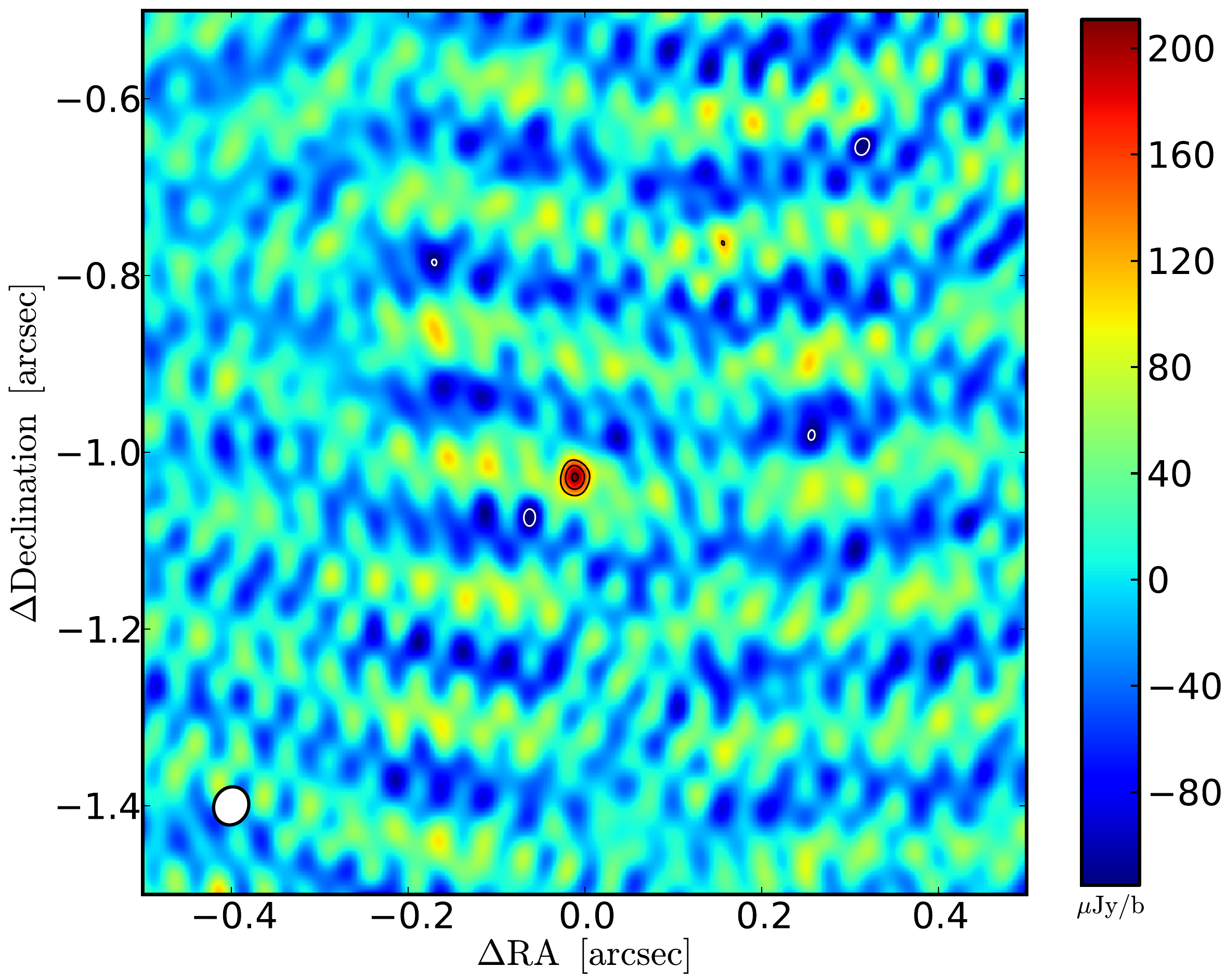}
\caption{EVN 1.66 GHz continuum map of IRAS~10214. Over-plotted are contours starting from -4, 4-$\sigma$ and increasing in steps of 2-$\sigma$, where $\sigma$ = 23~$\mu$Jy per $44\, \times \,39$ mas$^2$ beam (white ellipse at bottom left, position angle = -25.1$^\circ$ east of north). The white contours indicate negative contours and black contours are positive. The peak and integrated flux densities are \mbox{$S_{\rm peak}  = 209 \pm 23~\mu$Jy} and $S_{\rm int}  = 220 \pm 37~\mu$Jy. The {\sl uv}-data are limited to a maximum of 5 M$\lambda$ based on the quality of the phase solutions derived from the phase calibrator. The observed noise structure results from the \evn synthesised beam. The map co-ordinates are centred on the lensing galaxy \hst\,F160W centroid (RA = 10$^{\rm h}$ 24$^{\rm m}$ 34.5622$^{\rm s}$, Dec = 47$^{\circ}$ 09' 10.809'', see {\bf D13a})}
\label{fig:evn}
\end{figure}

In Fig.~\ref{fig:evn} we show the \evn 1.66 GHz {\sc Cleaned} map with a \mbox{9-$\sigma$} detection. The peak flux density of the imaged data is $S_{\rm peak}  = 209 \pm 23~\mu$Jy. Fitting a Gaussian to the {\sc Cleaned} map returns an integrated flux density $S_{\rm int}~=~220~\pm~37~\mu$Jy, which suggests that the source is unresolved. The measured flux density and position of the detection was checked for a large number of subsets of the data, including splits in frequency, time, and antenna selection. All these tests yield consistent positions and a flux density peak that ranges from $S_{\rm peak} \sim 190-220\,\mu$Jy. The positional uncertainty is $\sigma_{\theta} = 2.2$~mas, which is defined by $\sigma_{\theta} = 0.5\,{\rm FWHM}/({\rm S/N})$ \citep{Condon1997}.

As shown in Fig.~\ref{fig:EVNHSTCO}, the \evn detection does not appear exactly co-incident with the \hst\,F814W (rest-frame ultraviolet) peaks or the \co total intensity map peak. First inspection shows that it lies roughly halfway along the main \hst\,F814W arc (in RA), and $>100$~mas northward of the centre of curvature. It is located $\sim$170~mas towards the north-west of the main (eastern) \hst\,F814W peak. As discussed in {\bf D13a}, the \hst astrometry is in agreement with determinations from \citet{Nguyen1999,Eisenhardt1996,Evans1999} and Simpson~et~al.~(in preparation) to within $\Delta \theta <$ 10 mas.

\subsection*{Comparison with \hst image-plane configuration}

The lensing configuration, detailed in {\bf D13a} and briefly reviewed in \S1, includes an arc at near-infrared wavelengths to the south of the lensing galaxy as well as a counter-image towards the north of the lens (see fig.~2, {\bf D13a}). Our \evn observations here do not detect any emission in the region of the \hst\,F814W and \hst\,F160W counter-images. We place a weak, 5-$\sigma$ limit on the counter-image flux density of $S_{\rm 1.7,min} \lesssim 100 \, \mu$Jy since VLBI detections are generally only considered robust if greater than 5-$\sigma$, due to poor {\rm uv}-coverage\footnote{Since our {\sl uv}-coverage is comparatively good for a VLBI observation, 5-$\sigma$ is a relatively conservative upper limit.}. We place the same detection limit for any potential emission that is co-located with the secondary \co peak (labelled `A' in {\bf D13b}), and for the lensing galaxies identified in {\bf D13a}~(Fig.~9). Note that these limits were obtained by shifting the phase centre to the \hst F160W centroid in each case.

As stated in \S1, one of the questions raised in the past is whether or not the \hst\,F814W structure along the arc is intrinsic or if there are multiple images that are convolved together by the \hst PSF. The \hst\,F814W angular resolution was insufficient to unambiguously determine this, however, the higher angular resolution afforded by the EVN will determine if there is a radio core in the proximity of the \hst\,F814W emission and more specifically if this radio core has multiple images within the \hst\,F814W image plane arc. These VLBI observations will therefore provide further constraints on the lens model, should a detectable radio core be present.  

\begin{figure*}
\centering
\includegraphics[width=0.97\textwidth]{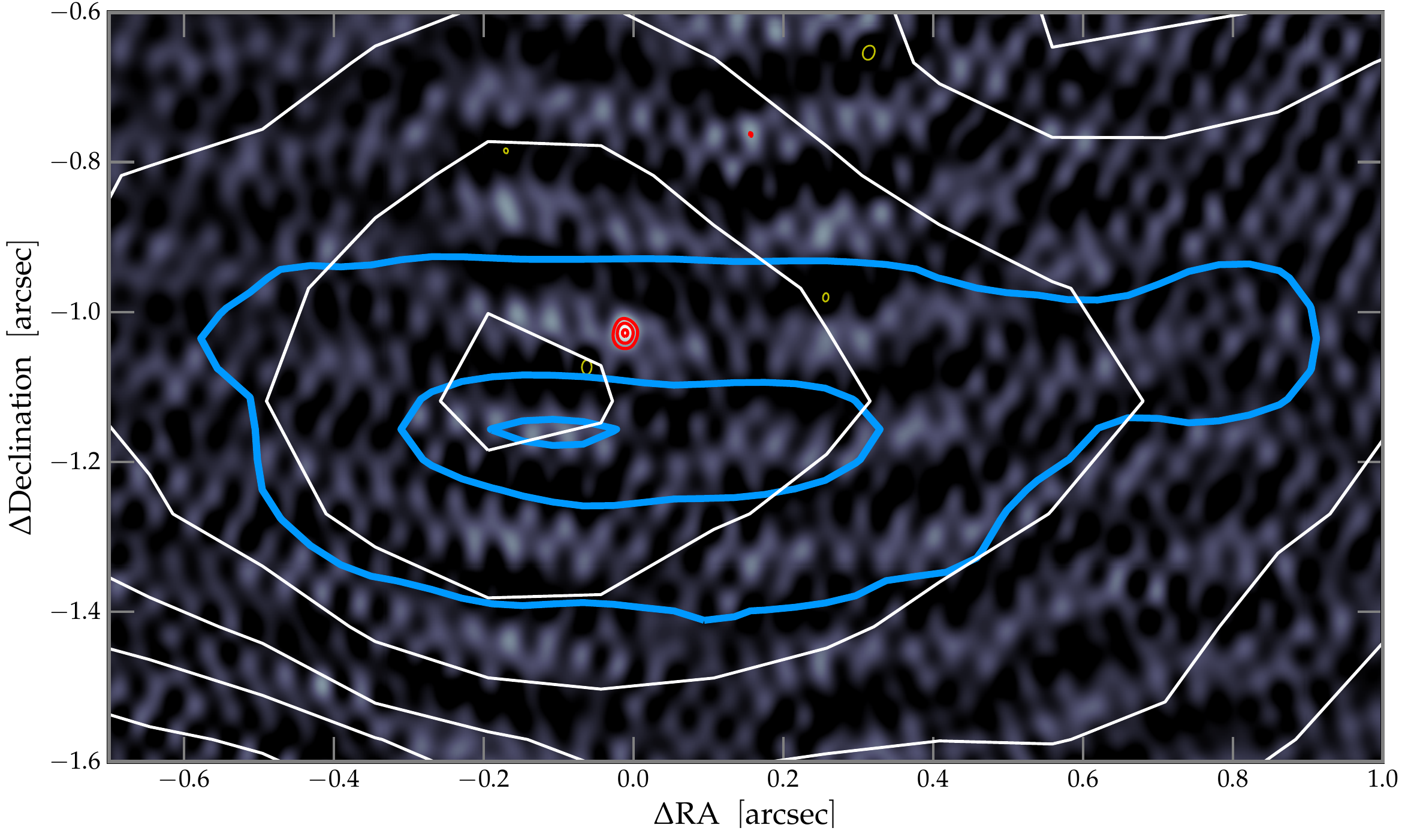}
\caption{Zoomed-in view of the \evn 1.66 GHz continuum map from Fig.~\ref{fig:evn} with red \emph{positive} contour levels at 4, 6, 8-$\sigma$ and yellow \emph{negative} contours at -4-$\sigma$. The \co total intensity map is over-plotted in white contours with 2-$\sigma$ intervals that start at -3, 3-$\sigma$ ($\sigma_{\rm CO}$ = 60~$\mu$Jy\,beam$^{-1}$). The blue contours show the \hst\,F814W (rest-frame UV) arc at 2, 50, 95 percent of the peak flux density. The \co and \hst\,F814W point spread functions are $\sim 0.7$ and 0.1 arcsec respectively. No additional \evn images appear to be present within the extent of the \hst\,F814W arc.}
\label{fig:EVNHSTCO}
\end{figure*}

The relative magnification of these three split images would be defined by the cusp relation \citep[e.g.][]{Blandford1989,Keeton2003}, which states that the flux density of the two outer images should have a sum equal to the flux density of the central image. Therefore, if two outer images were present, they should \emph{each} have a flux density of $S_{\rm outer} \sim 105 \, \mu$Jy, equivalent to a \mbox{4.5-$\sigma$} detection in the \evn map. No evidence of additional images is present within the extent of the \hst optical/NIR arc, as illustrated in Fig.~\ref{fig:EVNHSTCO} which shows the \evn map with overlaid \hst\,F814W arc contours. The cusp relation does assume a smooth dark matter profile, however, substructure does not appear to impact image magnification ratios substantially \citep[e.g.][]{McKean2007,More2009}. Therefore, we assume the \evn image-plane is comprised of a main image (detected here at 9-$\sigma$) and an undetected counter-image that is in the vicinity of the \hst detected counter-images.

\section{Ultraviolet Polarisation Properties}\label{sec:pol}

Polarisation is a potential consistency check of a lens model, since theoretical arguments show that the polarisation angle undergoes a negligible change in galaxy-scale lensing \citep[i.e. in the weak field limit,][]{Dyer1992}. Since scattered light has a polarisation angle perpendicular to the vector between the source of the photons and the scattering clouds, we would therefore expect the vector between the AGN and scattering cloud(s) to have a position angle  $\phi_{\rm pol} - 90^\circ$, where $\phi_{\rm pol}$ is the ultraviolet/optical polarisation angle. In this section we collate previous rest-frame ultraviolet polarimetry measurements and review the spatially-resolved \hst imaging polarimetry reported in \citet{Nguyen1999}, before comparing the polarisation with the derived source-plane properties of the \evn detection in \S\ref{sec:srcplaneVLBI}.

A number of polarisation observations have been performed towards IRAS~10214 \citep[e.g.][]{Lawrence1993,Goodrich1996,Nguyen1999}, all of which report high polarisation in the observed optical frame. A review of these results is summarised in Table~\ref{tab:pol} which shows that the shorter wavelengths have higher polarisation percentages. From the three independent polarisation angles listed in Table~\ref{tab:pol}, we calculate the uncertainty-weighted mean polarisation angle $\bar \phi_{\rm pol} = 68.9 \pm 1.3$ which we use in our analysis of the source plane in \S\ref{sec:srcplaneVLBI}.

\begin{table}
\centering
\scriptsize
\begin{tabular}{c  c  c  c  c   }
\hline
{\bf Telescope} &  {\bf Bandpass}  &  {\bf Polarisation}  &  {\bf Position Angle}  &  {\bf Reference}  \\ 
                        &   nm                   &       percent                 &     degrees                 &         \\
 \hline   
{\sl Keck}            &   412-420            &     $26.5 \pm1.7$            &    $69.9 \pm 0.2$          &     \citet{Goodrich1996}               \\
{\sl Keck}            &   750-800            &     $15.9  \pm 0.4$           &    $69.9 \pm 0.2$          &      \citet{Goodrich1996}               \\
{\sl WHT}            &   400-1000          &     $16.4  \pm 1.8$           &    $75 \pm 3$          &      \citet{Lawrence1993}                \\
\hst                     &    F437M             &      $28 \pm 3$                  &    $62 \pm 2.7$          &      \citet{Nguyen1999}              \\

\hline
\end{tabular}
   \caption{Flux-weighted mean polarisation properties of IRAS~10214 from the literature. }
   \label{tab:pol}
\end{table}

\subsection*{Polarisation angle range}

The polarisation angle range reported in \citet{Nguyen1999} is $\Delta \theta_{\rm max} \sim 100^\circ - 65^\circ = 35^\circ \pm 5^\circ$. A simplified argument is made in \citet{Nguyen1999} that assuming the source of the photons is a point source, and the diameter of the scattering region ($D$) is known, then the projected distance between the scattering clouds and the photon source can be approximated by the formula $R~\approx~D~\theta_{\rm max}$. They assumed magnifications of $\mu_{\rm UV}$ = 45 to 250 to derive an intrinsic UV source diameter of $D$ = 40 to 100~pc (presumably by dividing the apparent source solid angle by $\mu_{\rm UV}$). This results in an AGN projected distance estimate of $R$ = 160 to 65~pc. These magnification estimates were based on the constrained range of arc to counter-image flux ratios (and magnification by proxy). Using the Markov Chain Monte Carlo (MCMC) derived \hst\,F814W source-plane diameter of $D = 2\,r_s = 720 \pm 180$~pc derived in {\bf D13a}, Sec.~4.1, we find a projected distance to the AGN of $R = 505 \pm 148$~pc. We note however, that this is not a direct comparison since the $\Delta \theta_{\rm max}$ measurement is made in the \hst\,F437M filter, while we have used the \hst\,F814W radius. We argue that this is a reasonable approximation, since the arcs in both these filters have very similar lengths ($\sim$0.7~arcsec) and unresolved widths. However, the \hst\,F814W size is possibly an overestimate since the polarised emission is likely to be biased towards the brightest, and therefore more compact emission. Evidence for this claim can been seen in Figure~2 of \citet{Nguyen1999}, which shows the extent of the polarised arc to be $\sim$0.55~arcsec long, after binning the 0.014~arcsec pixels into 0.042$\times$0.14~arcsec bins. We also note that a more accurate distance may be achieved with a more complex \hst\,F814W source model (rather than the circular Gaussian assumed in {\bf D13a}, particularly given the fact that the arc is unresolved in the north-south direction.

The polarisation angle range also provides another important piece of information.  In principle, it enables a robust estimate of the ionisation cone opening angle ($\phi_{\rm cone} = 0.5 \, \Delta \theta_{\rm max} = 17.5^\circ \pm 5^\circ$) assuming the incident photons originate from a point source. The covering factor can therefore be calculated $CF = 0.5 \, \Delta \theta_{\rm max} / (\pi/2) = 0.195\pm0.028$ by spherical symmetry. We use this observational constraint in our bolometric luminosity estimates in \S\ref{sec:lbol}.

\section{Source Plane}\label{sec:srcplaneVLBI}

\begin{figure*}
\centering
\includegraphics[width=0.97\textwidth]{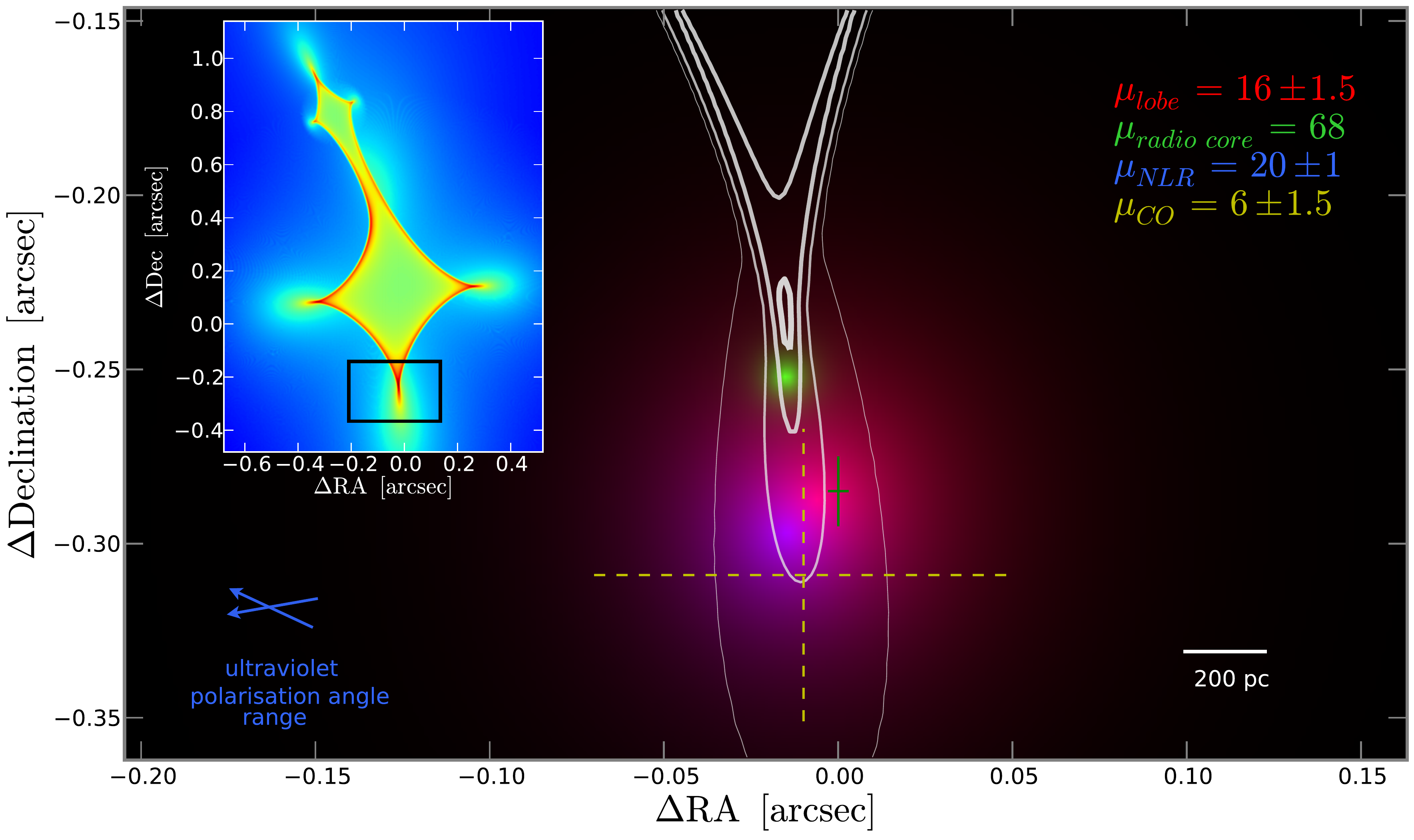}
\caption{Source plane reconstruction of IRAS~10214 showing the radio core (green, \evn 1.7~GHz), scattered quasar light (blue, \hst\,F814W) and radio lobe (red, \merlin 1.7~GHz). The yellow, dashed cross shows the \co total intensity map source-plane centroid and its uncertainty (the extent of the source-plane \co component is beyond this entire frame). The green cross indicates the 8~GHz map source-plane centroid and its uncertainty. The white contours represent lines of equal magnification extending from the caustic at levels $\mu$ = 10, 20, 50, 100. The blue arrows (bottom left) depict the range of UV polarisation angles along the \hst F437M arc. The inset shows the full lens caustic with colour-scale representing magnification and the black rectangle showing the borders of the enlarged region. }
\label{fig:srcplane}
\end{figure*}

We calculate the source-plane position of the \evn detection using the {\sc glamroc}\footnote{Gravitational Lens Adaptive Mesh Raytracing of Catastrophes, see http://kipac.stanford.edu/collab/research/lensing/glamroc} software. We use Lens Model A, as defined in {\bf D13a}. This assumes a Singular Isothermal Ellipsoid (SIE) potential for the main lens at $z = 0.893$, and a Singular Isothermal Sphere (SIS) potential for the secondary, line-of-sight galaxy at $z=0.782$. IRAS~10214 is assumed to be a point source, which is consistent with the Gaussian fit presented in \S\ref{sec:resultsVLBI}. The \evn source-plane co-ordinates are therefore the two remaining free parameters to solve for. We perform ray-tracing of a point source from the source plane to the image plane, and minimise the predicted image-plane position with respect to the measured \evn position. The resultant magnification of the best-fit point-source model is $\mu_{\rm EVN} = 68$. The magnification random error due to the $\sim2$~mas \evn positional uncertainty is of order 5~percent (derived by adding a 2~mas dispersion to the trial model position prior to ray-tracing), and therefore the systematic uncertainty of $\sim40$~percent  discussed in {\bf D13a}~(Sec.~3.5) dominates the absolute magnification uncertainty. The EVN 1.7~GHz component magnification is therefore $\mu_{\rm EVN} = 68 \pm 3 (\pm 27)$, where the systematic uncertainty is enclosed in parentheses. The \evn source-plane position is shown in Fig.~\ref{fig:srcplane}. 

Given the low probability that a small ($< 50$~mas) source is located within $<20$~mas from the cusp of our lens model, we describe a number of consistency checks on the \evn source-plane position, and its high magnification. Firstly, the \evn source is northward of the \hst UV/optical arcs as stated in \S\ref{sec:resultsVLBI}; and does not split into 3 images (at 4.5-$\sigma$ significance, see \S\ref{sec:resultsVLBI}); therefore it is most likely to the south of (or on) the cusp of the caustic. Secondly, the \evn detection is positioned roughly along the vector that extends from the centre of the lens and bisects the angle subtended by the rest-frame UV/optical arcs. This implies that the source-plane position must be very close ($<$ few milli-arcseconds) to the cusp of the caustic, since this is a region with large tangential deflection angles. The \evn point source is therefore essentially `wedged' between the inner-caustic and the \hst UV/optical arc, which will always result in high magnifications regardless of changes in the macroscopic lens model. In Fig.~\ref{fig:radmag}, we plot the predicted total magnification as a function of intrinsic source radius, assuming the source is centered on the \evn source plane position.

Assuming the \evn point source is the obscured active nucleus, then its source-plane position is  consistent with a number of multi-wavelength observations outlined below.

\begin{enumerate}


\item Simpson~et~al.~(in preparation) present two narrow-band maps centred on the redshifted \civ\ and \oiii\ lines, which are typical broad-line and narrow-line region emission lines respectively. They determine the centres of curvature of both and find that the \civ\ arc is roughly 50$\pm$20~mas northward of the \oiii\ centre of curvature for the highest S/N bins. At face value this implies that the BLR (or scattered BLR light) is northward of the NLR, which in our {\bf D13a} model is one-sided. This is consistent with resolved NLR imaging performed in the local Universe \citep{Liu1991,Simpson1997} and implies that the active nucleus is northward of the NLR and \hst\,F814W centers of curvature. The relative configuration of the emission associated with the BLR and NLR is therefore consistent with the source-plane \evn position, if it traces an obscured active nucleus. 

\item Comparison of the source-plane ultraviolet and radio core positions shows that the median ultraviolet polarisation angle ($\langle \phi_{\rm pol} \rangle = 82.5^\circ \pm5^\circ$, with a range $\sim 65^\circ -100^\circ$)  is roughly perpendicular to the vector connecting the \evn radio core and \hst\,F814W source-plane centroids ($\phi_{\rm core,uv} \sim 0^\circ$). This supports the view that the \evn detection is the dust-embedded source that emits the scattered, and therefore polarised, ultraviolet/optical radiation.

\item The lensing inversion predicts that the \evn and \hst\,F814W source-plane centroids are separated by a distance $R=370 \pm 148$~pc, where the uncertainty is dominated by the absolute magnification uncertainty. As calculated in Sec.~\ref{sec:pol}, the projected distance between the AGN and the scattered ultraviolet light is $R = 505 \pm 145$~pc, which is based on the ultraviolet polarisation angle range ($\Delta \theta_{\rm max} = 35\pm5^\circ$) and the predicted \hst\,F814W source-plane radius ($r_{\rm s} = 360 \pm 90$~pc). The two independently calculated projected distances between \evn and \hst\,F814W source-plane centroids are therefore consistent with one another.

The lensing inversion places the \evn-detected radio core at a position qualitatively consistent with the \citet{Nguyen1999} prediction of the AGN position. This is a promising example of the potential held by high sensitivity, high resolution spectro-polarimetry with next generation 30-40~metre optical/IR telescopes in accurately locating the obscured active nuclei in high-redshift galaxies.

\end{enumerate}



\section{Discussion}

\subsection{AGN Core or Highly-Magnified Star-Formation Clump?}\label{sec:agnSF}

At first glance, the \evn detection is likely to be the AGN radio core, however, careful consideration is required given the close proximity of the cusp of the caustic. IRAS~10214 was discovered at $\lambda_{\rm obs} = 100 \, \mu$m, a wavelength biased towards warm, partially AGN-heated ($T \sim 100$~K) dust. This is consistent with its AGN features across the spectrum and so the close proximity of an AGN core to the cusp would be expected. However, the probability that such a compact point source is within a few milli-arcseconds of the cusp is low. We must consider an alternative scenario where the \evn detection is co-spatial with a star-forming complex with an exceptionally large magnification ($\mu > 100$). This potentially has a higher probability, given the supernovae rate we expect from the large star formation rate (SFR) in IRAS~10214 ($SFR > 100\,{\rm M}_{\odot} \, {\rm yr}^{-1}$) and therefore the number of star-formation clumps.

We calculate the brightness temperature of the \evn-detected emission using Equ.~6 in {\bf D13a}. We assume a solid angle equal to the area of the beam effective radius (i.e. $\Omega_{\rm beam} = \pi \, \frac{1}{2} \theta_{\rm maj} \, \frac{1}{2} \theta_{\rm min}$ where $\theta_{\rm maj,min}$ are the FWHM values of the beam). The peak flux density is \mbox{$S(\nu_{\rm o}) = S_{\rm peak} = 209  \, \mu$}Jy, which results in a rest-frame 5.45~GHz brightness temperature lower limit of $T_{\rm B} > 5 \times 10^5$~K, since the source is unresolved. This lower limit seems to rule out a significant contribution from star formation, which appears to saturate at $T_{\rm SF}^{\rm max} \sim 10^5$~K \citep{Muxlow1994}. Note that these calculation do not depend on the magnification as the surface brightness, and hence the brightness temperature, are not affected by gravitational lensing. 
 
More distinguishing is the implied SFR density ($\Sigma_{\rm SFR}$) if we assume that the \evn-detected flux density is associated with star formation. Using Equ.~7 in {\bf D13a}, we calculate the implied SFR assuming all the emission is due to star formation. We assume a 1.7 GHz flux density $S_{\rm peak} = 209  \, \mu$Jy, a spectral index $\alpha = 0.8$ and set the intrinsic source size equal to the \evn 1.7~GHz beam. This results in a lower limit of $\Sigma_{\rm SFR} > 8.4 \times 10^4 \, {\rm M}_{\odot} \, {\rm yr}^{-1} \, {\rm kpc}^{-2}$. This is significantly larger than the theoretically motivated and observationally supported value of $\Sigma_{\rm SFR}^{\rm max} \sim 1000 \, {\rm M}_{\odot} \, {\rm yr}^{-1} \, {\rm kpc}^{-2}$ \citep[see ][]{Elmegreen1999,Thompson2005,Scoville1997,Downes1998}. These arguments, in addition to those presented in \S\ref{sec:srcplaneVLBI}, appear to support that this is an AGN core, which is what we assume for the remainder of the paper. As argued in \S\ref{sec:srcplaneVLBI}, this is consistent with a number of multi-wavelength and rest-frame ultraviolet polarisation properties.

\begin{figure}
\includegraphics[width=0.47\textwidth]{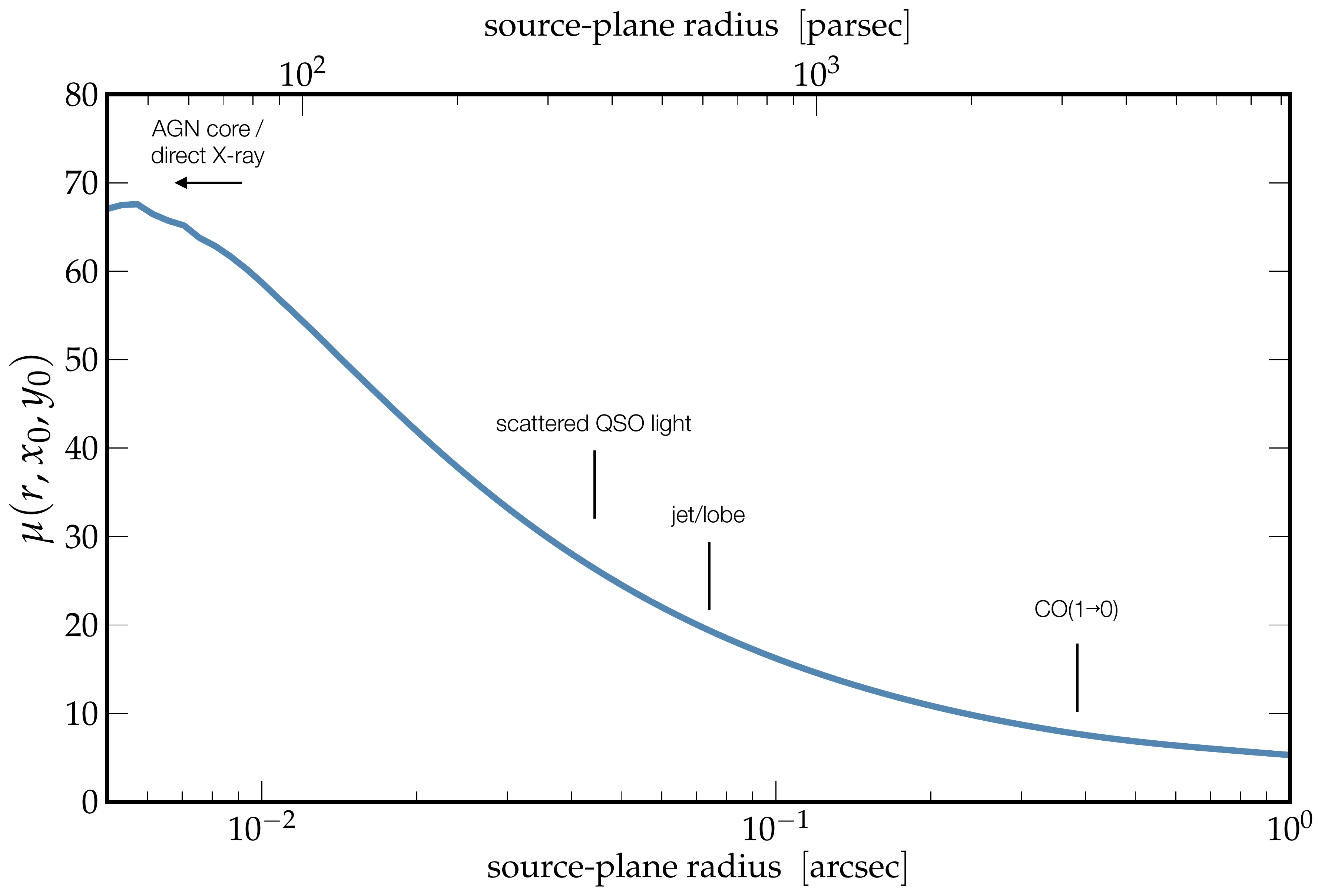}
\caption{ Total magnification as a function of source-plane scale radius, \emph{assuming the EVN source-plane centroid.} The annotations serve as a rough guide of the physical sizes. The centroids of the listed components are not the same as the \evn component and therefore the resultant magnifications are marginally different. The point source magnification at the best-fit \evn position is $\mu_{\rm EVN} = 68$. }
\label{fig:radmag}
\end{figure}

\subsection{Nature of \vla 8 GHz Map Peak}

\begin{figure}
\includegraphics[width=0.47\textwidth]{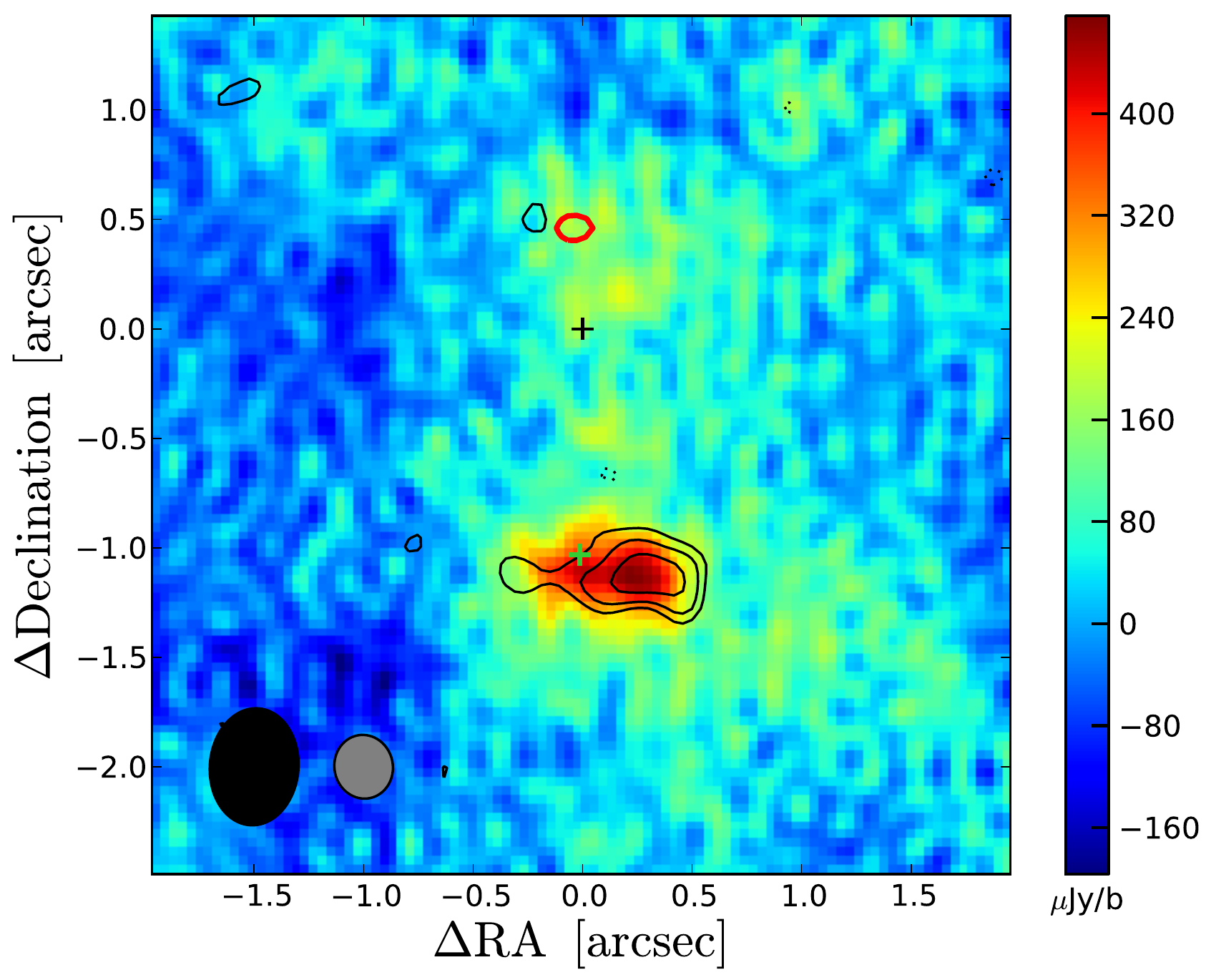}
\caption{  \merlin 1.7 GHz map shown in colour-scale with $\sigma \, \sim$ 46 $\mu$Jy per 405 $\times$ 349 mas$^2$ beam (position angle = -4.35$^\circ$). Over-plotted in black contours is the 8 GHz \vla map with $\sigma \, \sim \,$11 $\mu$Jy per 292 $\times$ 267 mas$^2$ beam (position angle = -79.64$^\circ$). Contours are at $\pm$3$\sigma$ and increase by a factor of $\sqrt{2}$. Dashed lines represent the negative contours. The \merlin and \vla beams are illustrated in the bottom left corner as black and grey ellipses respectively. The green cross indicates the location of the \evn detected radio core, where the cross length is a factor $\gtrsim 10$ larger than the \evn positional uncertainty. The red ellipse towards the north is the \hst\,F814W counter-image contour. The black cross indicates the Sersic-fitted centroid of the lensing galaxy as measured from the \hst\,F160W map}
\label{fig:overlay}
\end{figure}

In {\bf D13a}, we suggested that the \vla 8 GHz map could potentially be dominated by a radio core, based on the radio spectral index behavior from 330~MHz through 16~GHz. However, the 8 GHz centroid is significantly offset from the \evn detection presented here (see Fig.~\ref{fig:overlay}), with a separation of $\Delta \theta \sim 300 \pm 20 (\pm 200)$~mas, where the first uncertainty is the quadrature sum of the two random position uncertainties (8 GHz discussed in D13a), and the second uncertainty in parentheses is the systematic uncertainty based on the radio-optical reference frame alignment. The nature of the 8~GHz emission is puzzling, since we would expect it to be co-located with the \evn detection, especially since their flux densities are very similar ($S_{\rm 8GHz}  = 280 \, \mu$Jy), assuming a flat radio spectrum. It does not seem that this discrepancy can be discarded on the basis of insufficient astrometric accuracy. This is because the 8~GHz centroid was measured for two independent observations performed in 1991 and 1995, both with the \vla in A-configuration and hence a resolution of $\theta_{\rm syn} \sim 0.23$~arcsec. As discussed in \S\ref{sec:evnobs}, the astrometry of the \evn map appears secure to within $\lesssim 2$~mas. The true nature and extent of the 8~GHz emission will be further probed by a {\sl C-}band \jvla polarisation observation of IRAS~10214 in A-configuration ($\sim$30~mas angular resolution) with 2~GHz bandwidth. With the \evn map we can estimate a 5-$\sigma$ limit on the spectral index of the 8 GHz peak of $\alpha < -0.54$, assuming it is a point source and defining \begin{math} \alpha \equiv   - \log(S_1 / S_2) / \log( \nu_1 / \nu_2) \end{math}.

\subsection{Resolved Out Radio Emission}

The radio core flux is $\sim$20~percent of the total flux in the 1.7 GHz {\sl MERLIN} map presented in {\bf D13a}. If we assume the remaining flux is associated with star formation and a spectral index $\alpha = 0.8$, the implied star formation rate is $SFR \sim 3.2 \times 10^4 \,\mu^{-1} \,  {\rm M}_{\odot} \, {\rm yr}^{-1}$ following Equ.~7 in {\bf D13a}. If the \merlin 1.7~GHz magnification ($\mu_{\rm 1.7} = 16 \pm1.5$) derived in {\bf D13a} is assumed to be equivalent to the star formation magnification, then this results in an intrinsic star formation rate of $SFR \sim 2 \times 10^3 \,  {\rm M}_{\odot} \, {\rm yr}^{-1}$. This is an exceptional star formation rate, which suggests that there may be significant flux in the radio jet/lobe components that are resolved out by the \evn observations. It is unlikely that the total magnification boost of the star formation emission region is higher, since any disk with a reasonable star formation size ($\gtrsim 1$~kpc) undergoes a magnification boost of $\mu \lesssim 10$.

\section{Quasar Bolometric Luminosity}\label{sec:lbol}

Having detected the radio core and made an estimate of its magnification, we now wish to use these results to estimate the quasar bolometric luminosity in IRAS~10214 using five methods in the X-ray, ultraviolet and mid-infrared


\subsection{\oiii\ Equivalent Width $L_{\rm bol}$ Estimate}\label{sec:oiii}

 The first method uses the equivalent width ($EW$) of the forbidden \oiii$\lambda$5007  line to measure the `missing' optical continuum flux as a result of the obscuring geometry \citep{Miller1992}. Their method measures the average \oiii\ equivalent width of 76 type-I quasars from the Bright Quasar Survey (BQS), which provides a way to estimate the obscured AGN continuum flux, and hence the extinction. The idea is that the narrow-line-region flux will be largely unaffected by the orientation of the quasar since it is above and beyond the putative torus, whereas the central AGN continuum would be obscured under the unified quasar model \citep{Antonucci1993}. Since the equivalent width measures the fraction of energy  of a spectral line relative to the underlying continuum, the greater the central AGN obscuration, the more exaggerated the equivalent width of a narrow line like \oiii\ will be from its intrinsic value. 

\citet{Miller1992} measure an average rest-frame \oiii\ equivalent width ${\langle EW \rangle}_{\rm OIII}$~=~$24 \pm_{12}^{25} \AA$ for their BQS survey sample. \citet{Serjeant1998} measure the \oiii\ rest-frame $EW$ of IRAS~10214 to be $EW_{\rm OIII}$~=~580~\,(1+z)$^{-1}$~=~176~$\AA$. Two approaches can now be followed to estimate the bolometric luminosity. The first simply assumes that given a measurement of the \oiii\ line flux density, one can estimate the bolometric luminosity with the relation 

\bea
L_{\rm bol,OIII^1} \ = \ L_{\rm OIII}\, \mu_{\rm NLR}^{-1} \ BC ,
\eea

\noindent where $L_{\rm OIII} = 2.05 \times 10^{37}$ W \citep{Serjeant1998,Lacy1998}, $\mu_{\rm NLR} = 20 \pm 1$ is the NLR magnification derived in {\bf D13a} from the \hst\,F814W map, $BC = 89.9$ is a bolometric correction factor calculated from the average quasar \oiii\ equivalent width \citep{Miller1992} and a correction factor to convert the specific luminosity at 5100~$\AA$\ to a bolometric luminosity \citep{Elvis1994}. This results in a bolometric luminosity estimate of $L_{\rm bol,OIII^1} = 2.4 \pm{0.6} \times 10^{11}$~L$_{\odot}$.  
 
The second estimate, $L_{\rm bol,OIII^2}$, is less direct. It extrapolates an optical continuum flux based on a dust-reddened quasar SED fit to the ultraviolet flux data points. Scattered QSO light is presumed to dominate the latter. By comparison of the observed IRAS~10214 and average quasar \oiii\ equivalent widths, the argument can be made that only $13.6 \pm_{6.8}^{14}$~percent of the AGN core continuum emission is observed at 5100~$\AA$. This is calculated by taking the ratio of $EW_{\rm OIII}'$/${\langle EW \rangle}_{\rm OIII}$, where $EW_{\rm OIII}'$ is the measured \oiii\ equivalent width; ${\langle EW \rangle}_{\rm OIII}$ is the average \oiii\ equivalent width \citep{Miller1992}. Therefore, we can estimate the true specific luminosity at 5100 $\AA$ as follows:

\begin{eqnarray}\nonumber
L_{\rm 5100,true} & = & L_{\rm 5100,obs}' \frac{EW_{\rm OIII}'}{{\langle EW \rangle}_{\rm OIII}} \\ 
                 & = &  L_{\rm 5100,obs} \, \mu_{5100}^{-1} \ \frac{EW_{\rm OIII}}{{\langle EW \rangle}_{\rm OIII}} \, \frac{\mu_{5100}}{\mu_{NLR}}, 
\end{eqnarray}

\noindent where $L_{\rm 5100,obs}'$  is inferred from the lensed, host-galaxy-reddened \emph{quasar} flux density at 5100 $\AA$; $L'_{\rm 5100,obs}$ = 1.9~$ \times \, 10^{23}$~W\,Hz$^{-1}$; ${\mu_{5100}}/{\mu_{NLR}}$ accounts for the difference in magnification between the narrow-line region and the AGN nucleus, and $EW_{\rm OIII}$ is the equivalent width in the absence of differential lensing effects. Note that $\mu_{5100} = \mu_{\rm AGN}$ since we are interested in the specific luminosity of the quasar at 5100~$\AA$.

\begin{figure}
\includegraphics[width=0.48\textwidth]{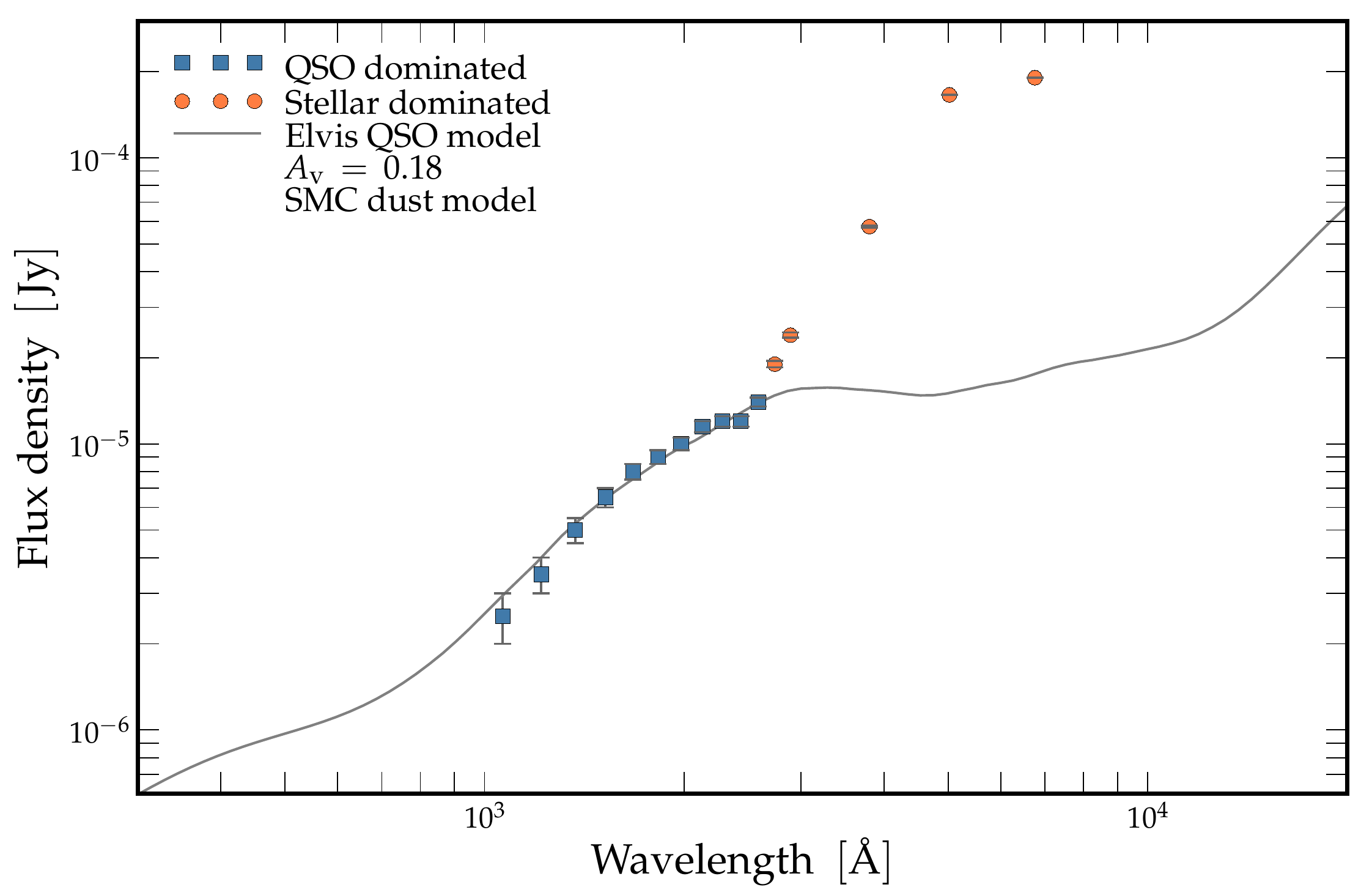} 
\caption{Ultraviolet fit to a reddened quasar model. The dust extinction model is from \citet{Pei1992} and the intrinsic quasar model is an average from the sample of \citet{Elvis1994}. Only the blue squares are included in the fit as these are presumed to be dominated by scattered quasar light, as suggested by the high degree of polarisation. We argue that orange circles are dominated by the host galaxy's stellar component which is supported by the lower magnification seen at rest-frame optical/NIR wavelengths (as surmised from lower arc to counter-image flux ratios).    }
\label{fig:uvelvisfit}
\end{figure}

To calculate  $L_{\rm 5100,obs}'$, we fit the average quasar spectrum \citep{Elvis1994} to the 11 UV photometric points between rest-frame 1000 -- 2750~$\AA$ and extrapolate to 5100~$\AA$ since this wavelength is dominated by stellar emission as discussed in {\bf D13a} and \S\ref{sec:BHbulgeratio}. This includes the \hst\,F814W photometric point and is therefore assumed to be scattered quasar emission given the high degree of polarisation. The UV photometric points are reddened by a range of dust models from \citet{Pei1992} to account for {\em host galaxy} extinction, since this is scattered QSO light. The fit results in a best-fit visual extinction $A_v$~=~0.18 with a Small Magellanic Cloud (SMC) dust grain model, with a $\chi^2_\nu$ = 1.3. The extinction is comparable to the $A_V \sim 0.3$ from \citet{Lacy1998} toward the NLR based on the He\,{\sc ii} 468.6 / 164.0 ratio of 4.3). The best fit extinction, bolometric luminosity, and $\chi^2_\nu$ have very little variance with subsets of the UV photometric points (from 4 to 11), however the goodness-of-fit begins to decrease substantially once the optical photometry is included, as one would expect with a significant stellar component. The full resolution UV spectrum from which these photometry points originate from, has a continuum baseline that \citet{Rowan-Robinson1993} fit with two linear models which meet at $\lambda_{\rm rest} \, \sim \, 2600 \, \AA$. This supports our selection of fitting boundaries, as well as the goodness of fit behaviour as this boundary is extended into optical wavelengths. The reddened quasar fit to the observed ultraviolet spectrum is plotted in Fig.~\ref{fig:uvelvisfit}.

This implies a bolometric luminosity $L_{\rm bol,OIII^2}$~=~$1.9 \pm_{1.0}^{2.0}$ $\times$ $10^{39}\,\mu_{\rm 5100}^{-1} \, {\mu_{\rm 5100}}/{\mu_{\rm NLR}} $~W, since $L_{\rm 5100,true} = L_{\rm OIII}/(24\pm^{25}_{12})$ as determined for unobscured quasars in \citet{Miller1992}. Assuming our previously derived magnification factors, this leads to a $L_{\rm bol, OIII^2}$~=~$2.8\pm^{2.8}_{1.4} \times 10^{11} \, {\rm L}_{\odot}$, which is 16~percent greater than the first $L_{\rm bol,OIII^1}$ method. We adopt the average of the two and note that the uncertainty is dominated by ${\langle EW \rangle}_{\rm OIII}$ scatter. The result is $L_{\rm bol, OIII}$~=~$2.6\pm^{2.6}_{1.3} \times 10^{11} \, {\rm L}_{\odot}$.

\subsection{Ultraviolet $L_{\rm bol}$ Estimate}

The scattered quasar light we observe is of course a fraction of the total quasar light. If we can determine the covering factor and appropriate magnification, then we can determine the intrinsic quasar bolometric luminosity. The magnification is reasonably assumed to be $\mu_{\rm NLR} = 20 \pm 1$ since the \hst\,F814W observation directly probes the emission assumed to be dominated by the scattered quasar light. The covering factor ($CF = 0.195 \pm0.028$ ) is approximated by the maximum change in polarisation angle ($\Delta \theta_{\rm max}$) of the scattered ultraviolet light as discussed in \S\ref{sec:pol}. The luminosity at 2400~$\AA$ (rest-frame \hst\,F814W) is $\nu  L_{2400} = 5 \pm 2 \times 10^{11} \, \mathrm{L}_{\odot}$. We also account for host galaxy extinction ($A_V$ = 0.18) based on the modeling described in \S\ref{sec:oiii}. This leads to a UV bolometric luminosity estimate of $L_{\rm bol,UV} = L_{2400} \, BC_{2400 \AA} \, 10^{A_V/2.5}\, /\mu_{\rm NLR}  = 2.7 \pm 1.1 \times 10^{11}\, f_{\rm scatt}^{-1} \, \mathrm{L}_{\odot} $. The scattering fraction $f_{\rm scatt}$, which is poorly constrained, will increase the derived value if much smaller than unity. We are fairly ignorant of the true $f_{\rm scatt}$ value, and plot the $L_{\rm bol,UV}$ estimate in Fig.~\ref{fig:xraypdf} (green line) assuming $f_{\rm scatt}$ = 100~percent.

\subsection{Mid-Infrared $L_{\rm bol}$ Estimates}

{\sl Spitzer} mid-infrared spectroscopy has proved a powerful technique to probe the heart of quasar host galaxies through dust heating and spectral features. In the case of IRAS~10214, the mid-IR spectrum is best fit with three components: (1) a typical star-forming dust temperature of 50 K; (2) a warm 210 K component; (3) a hot 600~K component \citep[see][]{Teplitz2006,Efstathiou2006}. We motivate appropriate covering factors and magnifications for the latter two components which we expect are predominately heated by the central AGN. This will allow additional estimates of the intrinsic bolometric luminosity.  

\subsubsection*{600 K Dust}

\citet{Efstathiou2006} fit a 600 K component to the mid-IR spectrum that has an apparent bolometric luminosity $L_{\rm bol,600K} = 7.4 \times 10^{11} \, \mathrm{L}_{\odot} \ CF_{\rm 600K}^{-1} \ \mu^{-1}_{\rm 600K}$. The covering factor ($CF_{\rm 600K}$) and magnification ($\mu^{-1}_{\rm 600K}$) at this temperature do not represent that of a distinct physical component, but rather a representative average of the hot dust component. We contend that this `component' is located within $\sim$50~pc of the radio core and therefore has a mean magnification of $\mu_{\rm 600K} \sim$68, as inferred from Fig.~\ref{fig:radmag}. The distance limit is based on mid-IR interferometric (using VLTI) size limits from nearby Seyfert galaxies \citep{Tristram2009,Tristram2011}. These authors find a size-luminosity relation $s = p \, L^{0.5}$, where $p = 1.8 \pm 0.3 \times 10^{-18}$\,[pc\,W$^{-0.5}$]. This implies a torus size of order $s= 30-55$~pc for IRAS~10214 for range of quasar bolometric luminosities determined for $L_{\rm bol,UV}$ and $L_{\rm bol, OIII}$. The covering factor for this component is set to the value derived in \citet{Mor2009}, who model the mid-IR spectra of 26 luminous QSOs with 3 dust components: (1) a clumpy torus; (2) a dusty narrow line region; (3) a hot blackbody representing the hottest dust around the nucleus. A 600 K dust component would clearly be associated with the last of these three components, for which the authors find a mean covering factor $\langle CF \rangle = 0.23\pm0.1$. This covering factor compares well with our adopted value $CF = 0.195\pm0.028$ which is based on the maximum change in polarisation angle. Based on the polarisation-determined covering factor and magnification we estimate a bolometric luminosity $L_{\rm bol,600K} = 4.8 \pm^{5}_{2.2} \times 10^{10} \, \mathrm{L}_{\odot}$.

\subsubsection*{210 K Dust}

This bolometric luminosity estimate follows the same reasoning as the 600 K component, however the average distance of clouds contributing to this temperature is greater than that of the hotter, 600~K dust and therefore likely to undergo a different amplification boost. The peak rest wavelength of this component is 14 $\mu$m, just longward of the 10 $\mu$m silicate feature. Based on evidence from silicate \emph{emission} in type II Seyferts \citep[e.g.][]{Sturm2006,Shi2006}, as well as resolved silicate imaging on $\sim$100 pc scales \citep{Schweitzer2008}, it appears that the $\sim$10 $\mu$m emission may be dominated by a combination of outer torus and NLR dust cloud emission. Following the multi-component mid-infrared modelling of \citet{Mor2009} who find a correlation between NLR cloud distance and bolometric luminosity, we estimate a NLR distance of $r_{\rm NLR} = 125 \pm^{35}_{25}$ pc based on the range of values derived for $L_{\rm bol,UV}$ and $L_{\rm bol, OIII}$. 

We make the simplified argument that the 210~K component has a 125~pc radius; the same covering factor as derived before ($CF = 0.195\pm0.028$); and is centred on the \evn point source which results in a magnification $\mu_{210\rm K} \sim 50$ (from Fig.~\ref{fig:radmag}). \citet{Efstathiou2006} and \citet{Teplitz2006} find a 210~K component bolometric luminosity of roughly $L_{\rm bol,210K} = 5.5 \times 10^{12} \, \mathrm{L}_{\odot} \ CF_{\rm 210K}^{-1} \ \mu^{-1}_{\rm 210K}$. We therefore find $L_{\rm bol,210K} = 5.6  \pm 3 \times 10^{11} \, \mathrm{L}_{\odot}$ from the assumptions above. This is higher than the other bolometric estimates, especially that based on the 600 K dust. This could be expected since some, potentially large, fraction of the 210 K dust heating will arise from the prodigious star formation in IRAS~10214. An estimate of this contribution is the fraction of luminosity a 80 K greybody (i.e. dust primarily heated by star formation; and roughly where the IRAS~10214 mid-IR spectrum peaks) contributes at $\lambda = 14~\mu$m, the wavelength that corresponds to the peak of a 210 K greybody. The 80~K component contributes 1.5~percent assuming equal peak luminosities for the two temperature components. Previous work has shown that the star formation dominates the bolometric luminosity of IRAS~10214 at the $>$90~percent level \citep[e.g.][]{Rowan-Robinson1991,Teplitz2006,Efstathiou2006}. If we assume a star formation to quasar $L_{\rm bol}$ ratio in the 90-95~percent range, then an 80 K greybody contributes $\sim$15-30~percent to the flux at 14~$\mu$m (210 K peak wavelength). This suggests the $L_{\rm bol,210K}$ estimate is significantly contaminated by star formation.

\subsection{X-ray $L_{\rm bol}$ Estimate}

This method uses an X-ray spectrum fit employing the models of Wilman \& Fabian~(1999) to find the best-fit column density of intervening hydrogen ($N_{\rm H}$). The X-ray models are Monte Carlo derived and assume an intrinsic unobscured type-I quasar X-ray spectrum of the form $L_{\nu}~\propto~\nu^{-0.9} \,e^{-\nu / \nu_{\rm c}}$ (where h$\nu_{\rm c}$ = 360 keV) as described in \citet{Madau1994}. \citet{Wilman1999} generated a library of models with a range of hydrogen column densities ($N_{\rm H}$), assuming 5 $\times$ solar metallicity as they find this provides a superior fit to the hard X-ray background which peaks at $\sim$30 keV. The models further assume a 2~percent flux component that is scattered off the central ionized medium. For simplicity, we do not consider models that include a reflection component off an accretion disk. In our MCMC routine we include a dispersion on the scattered percentage with FWHM = 10~percent (in the range 0 -- 100~percent). Although \citet{Wilman1999} assume a fixed scattering fraction, we wish to decrease the sensitivity to this assumption, particularly as they note that there is likely some dispersion in this quantity, as suggested by high infrared luminosity ($L \gtrsim 10^{13}$~L$_\odot$) IRAS galaxies observed with the ROSAT High Resolution Imager (HRI), that place upper limits on the $0.1\pm2.4$~keV fluxes and imply scattering fractions below $\sim$0.5 per cent \citep{Fabian1996,Wilman1998}. While the selection of a 10 percent dispersion FWHM for our MCMC chains is relatively arbitrary, it gives practically the same probability to the lower values for luminous infrared galaxies reported above, whilst also opening up the parameter space to incorporate the effect of higher values.

The resulting models are compared to the X-ray observations of IRAS~10214 with both Chandra and XMM-Newton Space Telescopes (which have consistent, albeit low S/N, results). Samples of the posterior probability distribution function (PDF) of the hydrogen column density $N_{\rm H}$ and bolometric luminosity $L_{\rm bol}$ are drawn in a MCMC algorithm similar to that described in {\bf D13a}. Fitting a library of X-ray models to the three X-ray data points is a weakly-constrained problem, and so an important caveat to these results is that we assume a hydrogen column density prior $N_{\rm H}$ with a mean of 10$^{24}$\,cm$^{-2}$ and a logarithmic FWHM of 2 dex. This is based on the evidence presented by \citet{Alexander2005} that IRAS~10214's nucleus must be Compton-thick ($N_{\rm H} > 1.5 \times 10^{24}\, {\rm cm}^{-2}$). They find the measured X-ray luminosity is 1-2 dex lower than expected given the measured \oiii\ line strength and the $L_{\rm [OIII]} - L_{\rm X}$ correlation \citep[e.g.][]{Netzer2006}. If the obscuring gas column density is significantly larger than $N_{\rm H}$ $\sim 10^{25}$~cm$^{-2}$ then our assumption breaks down and the bolometric luminosity derived here is likely to be too low. However, if one assumes the X-ray luminosity is dominated by the AGN (and not less-obscured star formation) then the $L_{\rm \sc OIII}$ shows that the hydrogen column density cannot be significantly larger than 10$^{25}$ cm$^{-2}$ \citep{Alexander2005}. Note that if the narrow line region (and hence \oiii$\lambda$5007 flux) has a lower magnification factor than the active nucleus, as predicted by our lens model, this will increase the $L_{\rm [OIII]} - L_{\rm X}$ discrepancy, yielding greater evidence that the active nucleus is Compton-thick.

Under these assumptions, we find $\log \left( N_{\rm H} \right) \, = \,{23.5 \pm^{0.4}_{0.2}} \,$cm$^{-2}$ from the MCMC spectral fit illustrated in Fig.~\ref{fig:xraypdf}. By comparison with the model type-I X-ray spectrum we can estimate the intrinsic X-ray luminosity of IRAS~10214 and therefore the intrinsic bolometric luminosity. We find $L_{\rm bol, Xray} \, = \, 1.3 \pm^{1.2}_{0.7} \times \, 10^{39} \,  \mu_{\rm X}^{-1}$ W (3.4 $\times$ 10$^{12}\, \mu_{\rm X}^{-1}$ ${\rm L}_{\odot}$), where $ \mu_{\rm X}$ is the magnification of the X-rays emanating from the AGN core. We do not have a strong constraint on the magnification of this unresolved X-ray detection given its low S/N and the absolute astrometric accuracy of Chandra (3$\sigma$ = 0.8~arcsec)\footnote{http://cxc.harvard.edu/cal/ASPECT/celmon/}, however, as an indicative estimate, we assume that $\mu_{\rm X}$ = $\mu_{\rm EVN}$, which yields an intrinsic bolometric estimate $L_{\rm bol, X} = 5.3 \pm^{5.1}_{3.0} \times$ 10$^{10}$ ${\rm L}_{\odot}$. We emphasise that this X-ray magnification could vary substantially from the assumed magnification since micro-lensing is observed in X-ray emitting regions of AGN \citep[e.g.][]{Pooley2007,Chen2012}, but not frequently in the radio emitting region \citep[e.g.][]{Koopmans2003}, suggesting that the former are constrained to smaller regions than the latter. However, co-spatial regions smaller than the VLBI core will tend to a magnification of $\mu_{\rm X} \sim \mu_{\rm EVN} \sim 68$. 

The 20-30 ks observations here combined with the magnification ($\mu \sim 68$) make this almost equivalent to the most sensitive X-ray field ever observed (Chandra Deep Field South, 4~Ms), however the lack of spectral resolution results in large degeneracies (particularly due to the scattering fraction) which are overcome to some extent with our assumed priors. Further progress will only be achieved with greater depth and spectral coverage/resolution.

\begin{figure}

\includegraphics[width=0.5\textwidth]{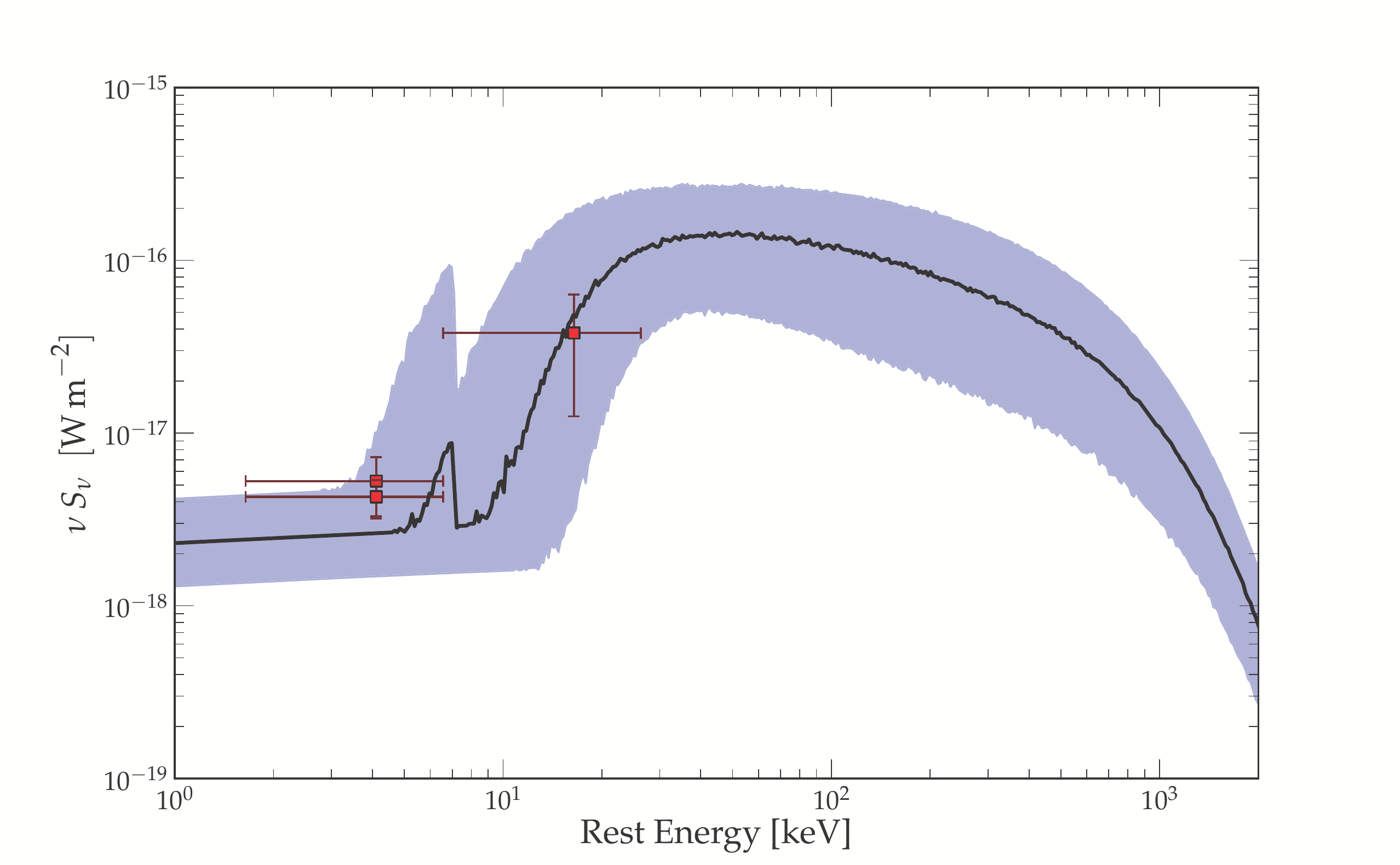} 
\caption{Best-fit X-ray model spectrum to the Chandra and XMM-Newton space telescopes. The shaded region corresponds the 68~percent confidence levels of the hydrogen column density ($N_{\rm H}$) and quasar bolometric luminosity ($L_{\rm bol}$). The bands correspond to 0.5-2 keV and 2-8 keV.  }
\label{fig:xrayfit}
\end{figure}

\subsection{Bolometric luminosity of IRAS~10214}

\begin{figure}
\includegraphics[width=0.47\textwidth]{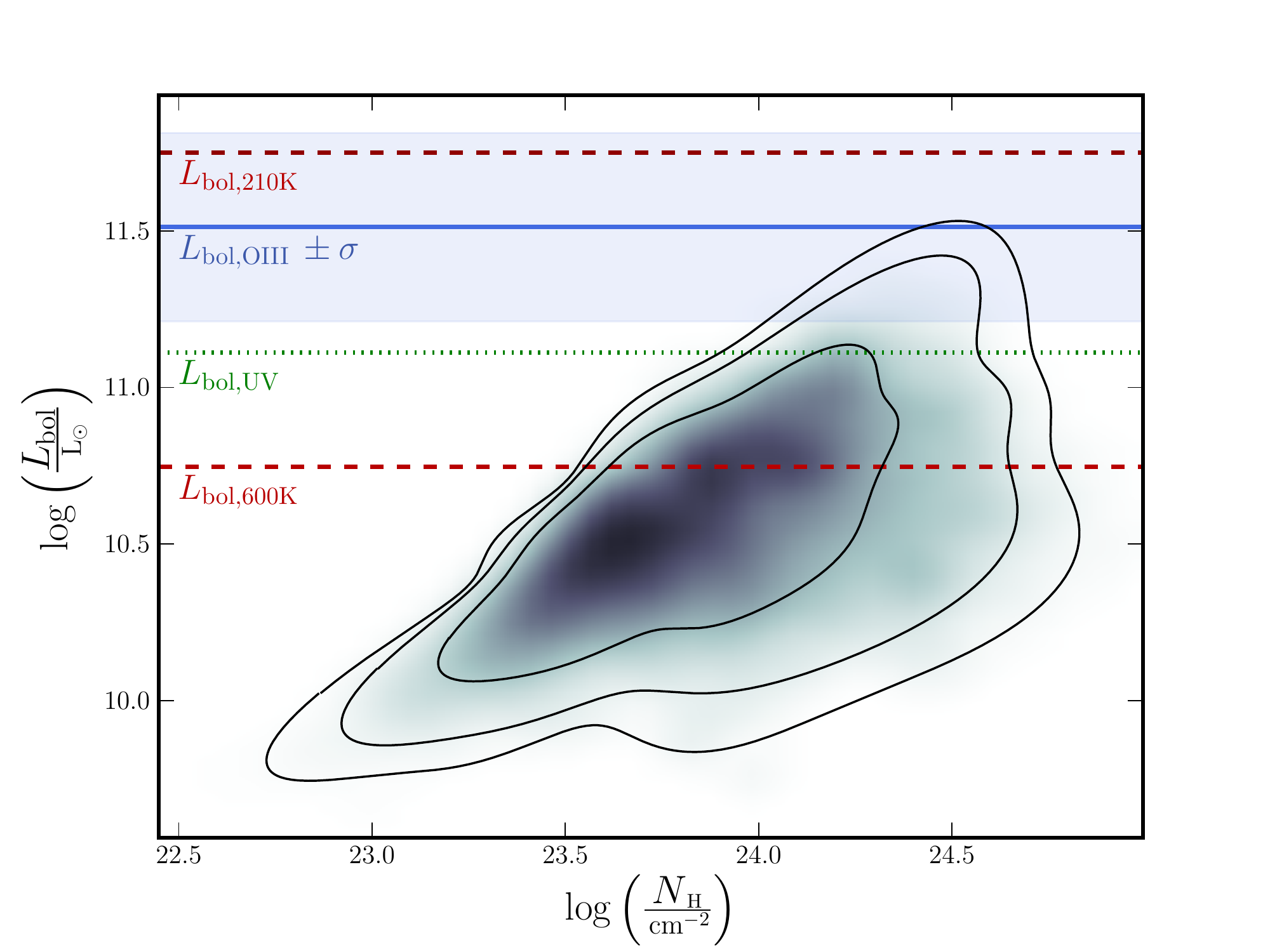} 
\caption{ Two dimensional posterior PDF assuming the \citet{Wilman1999} X-ray model spectra. The contours correspond to 68, 95 and 99~percent confidence levels. The blue line indicates the bolometric luminosity as derived from the \oiii\ equivalent width and the blue shaded horizontal band corresponds to the associated 1-$\sigma$ uncertainty. The two dashed red lines indicate the two dust bolometric luminosity estimates, while the green dotted line is the ultraviolet $L_{\rm bol}$ estimate. Note that the Compton limit is $\sim10^{24} \ {\rm cm}^{-2}$. For clarity, we have not included the $L_{\rm bol,UV}, L_{\rm bol,210K}$ and $L_{\rm bol,600K}$ uncertainties. They are dominated by the quadrature sum of the covering factor and magnification uncertainties and are of order 0.3-0.5 dex.}
\label{fig:xraypdf}
\end{figure}

In Fig.~\ref{fig:xraypdf} we show the 2D posterior PDF of the X-ray derived hydrogen column density ($N_{\rm H}$) and bolometric luminosity, which illustrates the degeneracy between these two parameters (below the Compton Limit). Over-plotted are two dashed red lines that indicate the two dust bolometric luminosity estimates. The blue horizontal line and shading indicate the $L_{\rm bol,OIII}$ estimate along with its 68~percent confidence levels. Finally, the green dashed line shows the estimate based on the scattered quasar light. While the uncertainty is large in all of the above methods ($\sim 0.3-0.5$ dex), they all show a reasonably consistent picture of a hidden quasar with intermediate luminosity. Furthermore, our highest estimate is the that which we expect to be most contaminated by star formation (i.e. the 210~K dust bolometric luminosity estimate). 

The five methods here are based on high-exposure space telescope observations and so it is challenging to see where we will make substantial progress in determining the bolometric luminosity of obscured quasars like IRAS~10214 before the next generation of telescopes or extended, dedicated programmes with current facilities. We adopt a quasar bolometric luminosity that is the weighted mean of the five methods presented here, which results in a value of $\log_{10}(\langle L_{\rm bol,QSO} \rangle / {\rm L}_\odot) = 11.34 \pm 0.27$, however, we stress the myriad systematic uncertainties associated with each method.

\section{Black Hole Parameters}

\subsection{Black Hole Mass} \label{sec:bhmass}

As discussed in \S1 and {\bf D13a}, \citet{Goodrich1996} found IRAS~10214 to have polarised broad emission lines (\ciii, \civ, \lya/\nv) with Keck spectro-polarimetry. Under the standard quasar unification model \citep{Antonucci1993}, this is consistent with a dusty toroidal structure obscuring a direct view of the AGN core and its associated broad-line region (BLR). However, a fraction of the BLR emission is scattered off dust and electrons into the observer's line of sight. This polarises the radiation providing a very useful tool in the dissection of AGN. Dust of course reddens the scattered radiation, whereas electrons do not alter the spectrum of the incident radiation with any wavelength dependence. Discerning the nature of the scatterers is possible in principle, however not at present in the case of IRAS~10214 due to the low S/N polarised spectrum.

 \citet{Goodrich1996} measured a polarised \civ\ line-width of $\Delta$v$_{\rm CIV}$ $\sim$ 6000 km\,s$^{-1}$. Following the \citet{Vestergaard2006} calibration, the black hole mass can be estimated from the \civ\ velocity width via:

\begin{eqnarray}\nonumber
\log \left( M_{\rm BH}({\rm CIV}) \right) & = & \log \left( \left[ \frac{ \Delta V_{\rm CIV}} {1000 \ {\rm km\,s}^{-1}} \right]^{\,2} \, \left[ \frac{\lambda L_{\lambda}(1350\, \AA)} {10^{44} \ {\rm ergs\,s}^{-1}} \right]^{\, 0.53} \right) \\
& &   + \ (6.66 \, \pm \, 0.01) 
\end{eqnarray}

We determine the value $\nu L_{1350}$ = 7.3 $\times$ 10$^{37}$ W based on the quasar bolometric luminosity derived in \S\ref{sec:lbol} and an average quasar spectrum \citep{Elvis1994}. This yields a black hole mass of \mbox{$\log_{10}(M_{\rm BH}/\mathrm{M_{\odot}})  =  8.36 \pm 0.56$}. Even with this large uncertainty, this places the black hole mass two orders of magnitude above the `knee' ($M^*$ value) of the $z= 0$ black hole mass function, however it is certainly not amongst the most massive black holes ($>$ 10$^9$  $\mathrm{M_{\odot}}$) observed at this epoch and earlier \citep[e.g.][]{Kurk2007}. Based on this black hole mass and the mean quasar bolometric luminosity derived in \S\ref{sec:lbol}, we can estimate the Eddington  Ratio,
\bea
\eta & = & \langle L_{\rm bol,QSO} \rangle / L_{\rm Edd}  \\ \nonumber
       & = &  2.2 \times 10^{11} \ { \rm L}_{\odot} / 7.4 \times 10^{12} \ {\rm L}_{\odot}    \\
       & = & 3^{+7}_{-2} \ {\rm percent.} \nonumber
\eea
\noindent This indicates (even with the large uncertainty) that the accretion rate is likely to be below the typical 10~percent level for actively accreting, radio-loud quasars, consistent with the low intrinsic radio core flux ($S_{\rm 1.7GHz} \sim 3 \, \mu$Jy).

 While there is some contention regarding the validity of the \civ\ FWHM as a black hole mass estimator \cite[e.g.][]{Baskin2005,Shen2008,Croom2011}, the polarised \civ\ line is perhaps the most robust black hole mass estimator we have at our disposal for this obscured quasar. We emphasise the large uncertainty of 0.56~dex as well as the caution raised by \citet{Baskin2005} regarding the use of the \civ\ FWHM in black hole mass estimates. Included in this uncertainty is the inclination dependency, since the broad-line region has been shown to be more disc-like rather than spherical \citep[see discussion in][and references therein]{Jarvis2002,Jarvis2006}. This is the first estimate of the black hole mass in IRAS~10214.

As a consistency check, we estimate the black hole mass, using a second, significantly less secure method based on the rest-frame 5~GHz luminosity and Eddington rate ($L/L_{\rm Edd}$, \citealt{Lacy2001}), based on their derived correlation between black hole mass and radio luminosity in the analysis of 60 quasars selected from the FIRST\footnote{Faint Images of the Radio SKy at Twenty cm, http://sundog.stsci.edu} Bright Quasar Survey \citep[FBQS;][]{Gregg1996,White2000}. Their correlation is re-arranged as follows:

\bea
\log_{10}(M_{\rm BH}) & = & \frac{\log_{10}(L_{\rm 5 GHz}) - 1.0 \, \log_{10} (L/L_{\rm Edd}) - 7.9}{1.9},
\label{equ:lacy2001}
\eea

\noindent where $M_{\rm BH}$ is in units of M$_{\odot}$ and $L_{\rm 5GHz}$ is in units of W\,Hz$^{-1}$\,sr$^{-1}$. The scatter in this relation is very large (1.1~dex), since a number of important physical effects are neglected such as black hole spin and Doppler-brightening \citep{Jarvis2002}.  Nonetheless, Equ.~\ref{equ:lacy2001} does provide an indicative black hole mass, which we calculate assuming the \evn 1.66~GHz (rest-frame 5.5~GHz) flux density of $S_{\rm int}  = 220 \pm 37~\mu$Jy, its derived magnification of $\mu_{\evn} = 68$, and accretion rates of $L/L_{\rm Edd} = $1, 10 and 100~percent. This results in black hole estimates of $\log_{10}(M_{\rm BH,5GHz} / \mathrm{M_{\odot}})  = 8.79, 8.27, 7.74 \pm 1.1$ for accretion efficiencies of $L/L_{\rm Edd} = $1, 10 and 100~percent respectively. The reasonable agreement is encouraging, however we use the \civ\ FWHM estimate for the remainder of the paper since this method is significantly more robust.

\subsection{Black Hole-Spheroid Mass Ratio}\label{sec:BHbulgeratio}

Based on the black hole mass estimate in \S\ref{sec:bhmass}, we now constrain the intrinsic stellar mass in IRAS~10214 in order to estimate the black hole to stellar mass ratio. We do not attempt a bulge-to-`disk' decomposition, since this requires detailed two-dimensional Sersic fitting of the source-plane inversion of a (preferably) pixel-based lensing inversion algorithm which is beyond the scope of this paper. We therefore make the assumption that the bulge mass dominates over the host stellar mass. The tentative 4000~$\AA$~break in \citet{Lacy1998} suggests that an older stellar population, more likely to be concentrated in a bulge, may dominate the rest-frame optical spectrum. The source-plane scale radius of the \hst\,F160W component in Deane~et~al.~(in preparation) is $r_{\rm s} \sim 500$~pc which is significantly smaller than the \co source-plane scale radius ($r_{\rm CO} = 5.7$~kpc) derived in {\bf D13b}. This is consistent with the expectation that a bulge component will have a radius significantly smaller than the extended gas reservoir. Finally, we assume that the stellar mass dominates the \emph{total} mass inside this loosely defined spheroid which is consistent with strong gravitational lensing results coupled with central lens stellar velocity dispersion measurements \citep{Koopmans2009,Treu2004}. The latter provide constraints on the slope of the total potential which when combined with the light distribution allow measurements of the stellar to total mass ratio at the bulge effective radius.

We select the \hst\,F160W map to estimate the total magnification since it has a significantly higher S/N than any of the other filters that are dominated by stellar emission. We derive a magnification of $\mu_{\rm stellar} = 14.1\pm^{1.6}_{1.2}$  (Deane~et~al.,in preparation), however, we re-emphasise the associated systematic uncertainty described in {\bf D13a} which is in the range of 20-40~percent. The \hst\,F160W lensing inversion is part of a detailed lensing analysis of several \hst filters which is beyond the scope of this paper. 

Spectrally-derived stellar mass estimates are challenging in the case of IRAS~10214 since the emission is AGN-dominated in the near-UV and the near-IR with an ill-defined transition into the stellar dominance at optical wavelengths, suggested by the tentative Balmer break. Ideally, we would make a mass estimate in the near-infrared, however, in addition to the AGN emission the lens galaxy makes a substantial, yet unconstrained contribution at this wavelength (Verma et al., in preparation). A compromise is to make a coarse estimate from a single filter that we believe is dominated by stellar continuum emission.

 We adopt the \hst\,F212N magnitude of $m_{\rm F212N} = 19.63 \pm 0.34$ determined by Simpson~et~al.~(in preparation) since this wavelength is a less biased tracer of stellar mass and it is the only \hst filter free of line contamination (Simpson~et~al., in preparation). The main difference between this and the \hst\,F160W map is likely to be dust extinction (between rest frame 490~nm and 650~nm) and therefore does not result in a substantially different solid angle (and hence the resulting magnification) between these two filters. We compare this apparent magnitude with the range of K-corrected \hst\,F212N apparent magnitudes derived from a wide range of star formation histories (SFH) using the models of \citet{Bruzual2003}. We vary the formation redshift between $z_{\rm f} = 6 -10$; metallicity between $Z = 0.008 - 0.05$; SFH models include SSP, singular burst and exponential, where the latter have $\tau$ values between 0.1 - 5 Gyr. Assuming the \hst\,F160W magnification $\mu_{\rm F160W} = 14.1 \pm^{1.6}_{1.2}$, this results in a range of predicted stellar masses of $\log(M^\star(z=2.3)/{\rm M}_\odot)  = 9.6 - 10.6$, and a mean of $ \langle \log(M^\star (z=2.3) / {\rm M}_\odot) \rangle =  10.1 \pm 0.5$, where the uncertainties are conservatively assumed to be half the full range.

Our dynamical and gas mass estimates in {\bf D13b}, imply a stellar mass of $M_{\rm stellar,dyn} = 6.8 \pm 1.7 \times 10^9$~M$_\odot$ which is calculated by assuming the mean gas fraction for a large sample of infrared luminous galaxies \citep{Bothwell2013} and a dark matter fraction within the \co effective radius \citep{Daddi2010}. The {\bf D13b} stellar mass is in good agreement with our \hst\,F212N estimate. The stated assumptions, along with this rough stellar mass of $\langle \log(M^\star (z=2.23) / {\rm M}_\odot) \rangle =  10.1 \pm 0.5$, enable an estimate of the central super-massive black hole to spheroid mass ratio, $M_{\rm BH}/ M_{\rm spheroid} =  0.02$, admittedly with large uncertainty ($\sim 0.8$~dex). However, this is $\sim$1.2~dex larger than the typical ratio found at $z \simeq 0$, however both are consistent with the suggested evolution of this relation for AGN host galaxies. For example, assuming $\alpha \sim 2 \pm 1$ (where $\alpha \propto M_{\rm BH}/M_{\rm spheroid}$), we expect an increase in the $M_{\rm BH}/ M_{\rm spheroid}$ relation by a factor of $11 \pm^{25}_8$ at a redshift of $z = 2.3$, in agreement with the our derived values.

The black hole to spheroid mass ratio is typically $M_{\rm BH}/ M_{\rm spheroid} \sim$0.1-0.2~percent at $z \sim 0$ \citep[e.g.][]{Merritt2001,Marconi2003,Haring2004}, but as stated above, are found to increase for high-redshift quasars, albeit with large uncertainty and selection effects, by a number of authors \citep[e.g.][]{Peng2006,McLure2006,Treu2007,Bennert2011}.  \citet{Peng2006} investigated a sample of 31 gravitationally-lensed and 20 non-lensed AGN and found that the $M_{\rm BH}/M_{\rm bulge}$ mass increased by a factor of $\gtrsim 4 \pm^{2}_{1}$ for $z> 1.7$. \citet{McLure2006} performed a similar study with the 3CRR sample \citep{Laing1983}, which consisted of 170 radio-loud, low-frequency selected AGN. They found an evolution of  $M_{\rm BH}/M_{\rm spheroid} \propto (1 + z)^\alpha$ where $\alpha = 2.07 \pm 0.76$. \citet{Treu2007} and \citet{Bennert2011} find a consistent values of $\alpha = 1.5 \pm 1.0;  1.96 \pm 0.55$ for samples with mean redshifts of $z \sim 0.36$ and $z \sim 2$ respectively. The former sample was made up of 20 Seyferts, while the latter comprised of 11 X-ray selected, broad-line AGN in a GOODS-N/S sample. There are clearly a number of independent studies that suggest quasar black hole growth precedes the full development of the stellar component to some degree, however, selection biases could yield similar results, as argued in \citet{Lauer2007}.

Although the uncertainties are large, the overall picture strongly supports that the majority of the black hole growth precedes the stellar mass assembly for this obscured quasar (assuming that IRAS~10214 will evolve to be consistent with the local  $M_{\rm BH}/ M_{\rm spheroid}$ relation). This is contrary to the `under-massive' black hole masses measured in $z \sim 2$ SMGs by \citet{Alexander2008} and \citet{Biggs2010}. \citet{Alexander2008} made $M_{\rm BH}$ estimates in 6 broad line SMGs using H$\beta$ and H$\alpha$ line-widths using the \citet{Greene2005} virial estimator, and found that (a) they have black hole masses $\gtrsim 3$ times smaller than `normal'  galaxies in the local Universe of comparable mass; and (b) they have $\gtrsim 10$ times smaller black holes masses than that predicted for $z\sim2$ populations. From their Fig.~4, they derive a $M_{\rm BH}/ M_{\rm spheroid}$ ratio $\approx 2.5 -4 \times 10^{-4}$ for the CO-dynamics and stellar mass determined bulge mass. \citet{Biggs2010} performed a VLBI survey of 6 SMGs with a mean redshift of $z \sim 2$. They calculate black hole masses (and upper limits) based on the \citet{Lacy2001} $L_{\rm 5GHz}$ method and their 1.6~GHz VLBI observations of all 6 SMGs. Although the uncertainty is very large (1.1~dex), all 6 of their objects have black hole masses (or upper limits) that are $\sim 0.2 - 1.3$~dex \emph{below} the expected value based on the \citet{Haring2004} finding that $M_{\rm BH}/ M_{\rm spheroid} = 0.0014 \pm 0.0004$ in the local Universe. The mean of their quoted stellar masses and black holes masses (as well as upper limits) implies a $\log(M_{\rm BH}/ M_{\rm spheroid}) \approx 0.0002$ with the quoted uncertainty of 1.1~dex.

Despite the large uncertainties, there appears to be 1-2 orders of magnitude difference between the black hole to spheroid mass ratio in SMGs and the obscured quasar IRAS~10214. This tentatively implies very different evolutionary paths for these two object classes, despite their apparent similarities in many parts of the global SED. However, this may be a result of the selection bias outlined in \citet{Lauer2007}, particularly since we demonstrate in this work that the AGN is preferentially magnified by an order of magnitude when compared to the spatially-resolved \jvla \co map in {\bf D13b}. 

Nonetheless, the high $M_{\rm BH}/ M_{\rm spheroid}$ ratio, relatively regular \co velocity map ({\bf D13b}), single \evn detection, highly magnified yet uniform \hst\,F160W morphology, are all consistent with IRAS~10214 being an example of a fairly typical AGN-host galaxy at $z \sim 2$, as found in \citet{Kocevski2012}. These authors present deep \hst\,F160W ({\sl H-}band) and \hst\,F125W ({\sl J-}band) observations of 72 intermediate X-ray luminosity ($L_{\rm X} \sim 10^{42-44}$~erg\,s$^{-1}$) AGN selected from the 4~Ms {\sl Chandra} observation of the Chandra Deep Field South, as part of the CANDELS survey \citep{Grogin2011}. They find that AGN host galaxies at $z \sim 2$ appear no more likely to be part of a major merger than a control sample of galaxies in the same mass range. One caveat, however, is that most major merger traits may be erased by the time the resultant AGN activity begins. The \citet{Kocevski2012} result is consistent with \citet{Schawinski2012} who show that a sample of $z\sim2$ highly obscured quasar host galaxies are not major mergers, but have reasonably smooth morphologies with Sersic indices consistent with disks. Furthermore, \citet{Cisternas2011} and \citet{Georgakakis2009} find consistent results at $z \sim 1$. These results form part of a growing consensus that major mergers do not play the dominant role in triggering intermediate luminosity AGN activity as outlined in the classical \citet{Sanders1988} scenario, where the merger-induced loss of angular momentum leads to black hole accretion. Moreover, these studies all find that AGN-hosts at $z \sim 1-3$ are predominantly disk-like in morphology, which is what the \co kinematics tentatively suggest for IRAS~10214 ({\bf D13b}).  

This paper (together with the series it forms part of) therefore suggests that IRAS~10214 hosts a super-massive black hole that has primarily grown in mass due to secular evolution, minor interactions and/or internal instabilities, rather than the classical major merger-induced scenario. We propose that IRAS~10214 is therefore a relatively typical intermediate luminosity AGN at $z \sim 2$, however its active nucleus is preferentially-lensed by an order of magnitude with respect to the star forming disk.

\section{Conclusions}

We have performed a deep ($\sigma = 23\,\mu$Jy\,beam$^{-1}$), 1.7 GHz \evn observation on IRAS~10214, a lensed $z=2.3$ obscured quasar with prodigious star formation. The VLBI observation provides a brightness temperature filter which is unobscured by dust and therefore allows us to image the obscured active nucleus with an effective angular resolution of $\lesssim$~50~pc at $z = 2.3$, after correcting for gravitational lensing. These \evn observations permit a number of conclusions to be drawn:

\begin{enumerate}

\item The AGN core as traced by the \evn 1.7 GHz detection has a flux density peak of $S_{\rm peak} = 209 \pm 23 \, \mu$Jy\,beam$^{-1}$ which makes up $\sim$20~percent of the total 1.7 GHz flux density of a {\sl MERLIN} map presented in {\bf D13a}. The \evn detection appears unresolved, with an integrated flux density of $S_{\rm peak} = 220 \pm 37 \, \mu$Jy. The remainder of the \merlin~1.7~GHz flux is likely to be split between star formation and large-scale ($\gtrsim$200 pc) jets. 

\item The fact that the AGN core is a single detection and is northward of the \hst\,F814W arc, strongly advocates that the \hst\,F814W arc is not a triple-image system that is merged by the \hst PSF. This supports the lens model derived by {\bf D13a} which suggests that the \hst\,F814W map is comprised of a single arc and the observed double peak corresponds to intrinsic structure at rest-frame ultraviolet wavelengths.

\item The AGN core is located at a position qualitatively consistent with the spatially-resolved polarisation properties of the ultra-violet map reported in \citet{Nguyen1999}. Moreover, it is consistent with their prediction of the active nucleus position based on these polarisation properties as well as the narrowband \hst observations which suggest that the BLR centre-of-curvature is northward of the NLR centre-of-curvature (Simpson~et~al., in preparation). 

\item If the AGN core position and size are well approximated by the radio core detected in this \evn observation, then the AGN is preferentially magnified by over an order of magnitude when compared to a spatially resolved \jvla \co map ({\bf D13b}), where the latter can be used as a star formation proxy. This confirms the effective `chromaticity' of this strong-lens system (discussed in {\bf D13a, D13b}), which is caused by different emission regions undergoing differing magnification boosts due to their relative size and position with respect to the caustic. This effect may be particularly relevant in the case of far-infrared (FIR) bright, strongly-lensed galaxies discovered by the {\sl Herschel Space Observatory} \citep{Negrello2010}.


\item The configuration of the \evn core and \hst ultraviolet arc support our previous claim that this is an asymmetric (or one-sided) source where we observe the line-of-sight NLR, whilst the `counter-NLR' is obscured by a significant host dust reservoir, consistent with examples in in the nearby Universe \citep[e.g.][]{Liu1991,Simpson1997}.

\item We derive a black hole mass of $\log_{10}(M_{\rm BH}/\mathrm{M_{\odot}})  =  8.36 \pm 0.56$ and a bolometric luminosity $\log_{10}(\langle L_{\rm bol,QSO} \rangle/{\rm L}_{\odot}) = 11.34 \pm 0.27$~dex, which suggests a low accretion rate $\eta \sim   3\pm^7_2$~percent, albeit with significant uncertainty.   

\item Our crudely derived black hole to spheroid mass ratios are consistent with the suggested evolution of this parameter in AGN host galaxies \citep{Peng2006,McLure2006,Treu2007,Bennert2011}. The $M_{\rm BH}/ M_{\rm spheroid}$ ratio in IRAS~10214 is 1-2 orders of magnitude larger than that of $z \sim 2$ SMGs based on optical and radio analyses \citep{Alexander2008,Biggs2010}, providing a tentative suggestion that these two object classes may follow different evolutionary paths (i.e. secular rather than merger-induced black hole growth). However, this picture is sensitive to the large black hole mass uncertainties and selection effects \citep[e.g.][]{Lauer2007}.

\end{enumerate}

These results point toward a intermediate luminosity AGN, however our X-ray modelling find it to be a highly obscured nucleus. The mean hydrogen column density \mbox{$\overline{N_{\rm H}} \sim 10^{23.5} \, {\rm cm}^{-2}$} explains the excessive \oiii$\lambda$5007 luminosity in comparison with the derived X-ray luminosity \citep{Alexander2008}. 
This X-ray modelling predicts a strong Fe K line to be observed if the required sensitivity is achieved. The unobscured, intrinsic quasar bolometric luminosity ($2.2 \times 10^{11} L_{\odot}$) is well supported with independent bolometric luminosity estimates and suggests inefficient, sub-Eddington accretion onto the central black hole which is consistent with the low intrinsic radio core flux ($S_{\rm 1.7GHz} \sim 3 \, \mu$Jy). Comparison with a large sample of AGN host galaxies at $z \sim 1.5 - 2.5$ \citep{Kocevski2012} suggests that IRAS~10214 is plausibly a fairly typical AGN host based on the lack of major merger traits and the relatively regular \co velocity map, consistent with a disk-like morphology. Therefore, the key difference in the case of IRAS~10214 may be that it has a preferentially magnified active nucleus. This case study is not only a clear demonstration of the significant SED distortion due to preferential lensing, but also a preview of what can be achieved with the enhanced sensitivity and new wide-field capabilities of the \vlba as well as the \evn which will detect large samples of low-luminosity AGN ($L_{\rm 5 GHz} \lesssim 10^{22}$~W\,Hz$^{-1}$\,sr$^{-1}$) over well-studied multi-wavelength fields \citep[e.g.][]{Middelberg2011}. These surveys will play an important role in characterising AGN activity over cosmic time, particularly for the obscured AGN population.

\section*{Acknowledgments}
This paper, as well as the series it forms a part of, is dedicated to the memory of Steve Rawlings. We are grateful to the \evn chair for awarding discretionary telescope time to perform a calibrator search in the vicinity of the target source - this was key to the success of the observation of IRAS~10214. We thank Zsolt Paragi who generously provided technical expertise in the planning and execution of the \evn observations. The European VLBI Network is a joint facility of European, Chinese, South African and other radio astronomy institutes funded by their national research councils. The National Radio Astronomy Observatory is a facility of the National Science Foundation operated under cooperative agreement by Associated Universities, Inc. This work made use of the Swinburne University of Technology software correlator, developed as part of the Australian Major National Research Facilities Programme and operated under licence. This effort/activity was supported by the European Community Framework Programme 6 and 7, Square Kilometre Array Design Studies (SKADS), contract no. 011938; and PrepSKA, grant agreement no.: 212243. RPD gratefully acknowledges funding that enabled visits to JIVE which were supported by the European Community Framework Programme 7, Advanced Radio Astronomy in Europe, grant agreement nl.: 227290.  PJM acknowledges support from the Royal Society in the form of a University Research Fellowship.


\label{lastpage}

\end{document}